\documentclass[useAMS,usenatbib]{mn2e}
\usepackage{graphicx}
\usepackage{pdflscape}
\begin{document}

\title[Sirius B gravitational redshift]{The gravitational redshift of Sirius B\thanks{Based on observations with the NASA/ESA {\it Hubble Space Telescope\/}
obtained at the Space Telescope Science Institute,  which is operated by the 
Association of
Universities for Research in Astronomy, Inc., under NASA contract NAS5-26555}}

\author[S.R.G.Joyce et al.]{S. R. G. Joyce$^{1}$\thanks{E-mail:
srgj1@le.ac.uk; sj328@le.ac.uk (SRGJ)}, M. A. Barstow$^{1}$\thanks{E-mail:
mab@le.ac.uk (MAB)}, J. B. Holberg$^{2}$, H. E. Bond$^{3,4}$, \newauthor S. L. Casewell$^{1}$, M. R. Burleigh$^{1}$ \\
$^{1}$Dept. of Physics \& Astronomy, Leicester Institute of Space and Earth Observation, University of Leicester, University Road,\\ Leicester, LE1 7RH \\
$^{2}$University of Arizona, LPL,Tucson, AZ, USA \\
$^{3}$Department of Astronomy \& Astrophysics, Pennsylvania State University, 
University Park, PA 16802, USA\\
$^{4}$Space Telescope Science Institute, 3700 San Martin Dr., Baltimore, MD 21218, USA}


\pagerange{\pageref{firstpage}--\pageref{lastpage}} \pubyear{2017}

\maketitle

\label{firstpage}

\begin{abstract}
Einstein's theory of General Relativity predicts that the light from stars will be gravitationally shifted to longer wavelengths. We previously used this effect to measure the mass of the white dwarf Sirius B from the wavelength shift observed in its H$\alpha$ line based on spectroscopic data from the Space Telescope Imaging Spectrograph 
(STIS) on the \textit{Hubble Space Telescope} (\textit{HST}), but found that the results did not agree with the dynamical mass determined from the visual-binary orbit. We have devised a new observing strategy using STIS where the shift is measured relative to the H$\alpha$ line of Sirius A rather than comparing it to a laboratory based rest wavelength. Sirius A was observed during the same orbit with \textit{HST}. This strategy circumvents the systematic uncertainties which have affected previous attempts to measure Sirius B. We measure a gravitational redshift of 80.65 $\pm$ 0.77 km s$^{-1}$. From the measured gravitational redshift and the known radius, we find a mass of 1.017 $\pm$ 0.025 M$_{\odot}$ which is in agreement with the dynamical mass and the predictions of a C/O white dwarf mass-radius relation with a precision of 2.5 per cent.

\end{abstract}


\section{Introduction}

In 1916, Einstein proposed three experimental tests of the theory of General Relativity \citep{Einstein_1916}. The first two tests were the deflection of starlight by the Sun, which could be observed during a solar eclipse, and the perihelion advance of the orbit of Mercury, which had been unexplained since it was discovered by Le Verrier in 1859. Both tests were found to be in accord with the new theory. The 3rd proposed test was to observe the gravitational redshift of the lines in stellar spectra. The effect is very small in main sequence stars ($\Delta\lambda $ $\sim$0.014 \AA ) making it difficult to measure, but it was realised that a compact star such as a white dwarf (WD) would produce a much larger effect. 

In 1924 Arthur Eddington, having just established his mass-luminosity relation for luminous stars, wrote to Walter S. Adams at Mt. Wilson Observatory inquiring about the possibility of measuring the ‘Einstein shift’ in Sirius B.  In his letter, Eddington estimated the shift to be approximately 28.5 km s$^{-1}$, and noted that the detection of a redshift velocity of this order would clearly demonstrate the ``incredible" density of Sirius B. Adams, aware that the ‘Einstein shift’ would be ‘large’, had already attempted to measure such a shift with the 100-inch Hooker telescope, and promised Eddington to make further attempts.  In the following year Adams published his result of 23 km s$^{-1}$ \citep{Adams_1925} and extended Eddington's original motivation for the observation to include a confirmation of the theory of General Relativity. Both Adams' observational estimate and Eddington's theoretical prediction of gravitational redshift for Sirius B were in error by nearly a factor of four, but this was not realised until many years later (See \citealt{Holberg_2010} for a detailed discussion of Adams-Eddington results).

The theory has since been tested to very high precision by terrestrial experiments, so the gravitational redshift effect has been employed as a means of measuring the $M/R$ ratio and hence the masses of WDs (e.g. \citealt{Greenstein_Trim_1967,Trimble_Greenstein_1972, Koester_1987,Wegner_1989,Oswalt_1991,Reid_1996, Holberg_2012,Parsons_2017}). The technique has the potential to provide high precision mass measurements of WDs which are invaluable as a test of the WD mass-radius relation \citep{Chandrasekhar1931}. 

Detailed tests of the mass-radius relation (MRR) require mass measurements with a precision of a few per cent in order to distinguish between models with different temperature and H-layer thickness. Such precision has been achieved for a small number of WDs in wide binaries using the dynamical method (e.g. \citealt{Bond_40Eri_17, Bond_sirius_17}) based on accurate orbit determinations. However, these results have required many decades of observations due to the long orbital periods. 

Another approach, the spectroscopic method \citep{Bergeron_eta92}, relies on fitting models to the broadened hydrogen lines visible in the spectra of DA type WDs. This technique is much easier to apply to large numbers of WDs as it only requires a single spectrum. Unfortunately, measurements of Sirius B using this method have not achieved the precision required for a definitive test of the MRR, as the uncertainty is $\sim$0.1 M$_{\odot}$ \citep{Joyce_2018_a}. 

The gravitational redshift method can potentially provide results with similar precision to the dynamical method but on much shorter time scales. Sirius B is the ideal test case because it already has a precisely determined dynamical mass \citep{Bond_sirius_17} which can be used as a benchmark to validate the results from the other two methods. It is also one of the few WDs which can be used to test the high mass end of the MRR, whereas most WDs have a mass closer to 0.6 M$_{\odot}$ (e.g. \citealt{Bergeron_eta92, Falcon_2010}). However, great care must be taken to obtain pure spectra which are uncontaminated by the light of Sirius A as was the case with the \cite{Adams_1925} data.

After the problems with the early Sirius B spectra were realised, \cite{Greenstein_1971} made a new estimate of the expected gravitational redshift, finding a much higher value of 83 $\pm$ 3 km s$^{-1}$. This was based on a radius of 0.0078 $\pm$ 0.0002 R$_{\odot}$ from fitting models to new spectra, combined with the dynamical mass of 1.02 M$_{\odot}$ \citep{van_den_Bos_1960}. The gravitational redshift of Sirius B was measured by \cite{Greenstein_1971} using the H$\alpha$ and H$\beta$ line and found to be in agreement with the revised predictions at 89 km s$^{-1}$. The precision on this measurement was limited to $\pm$ 16 km s$^{-1}$ due to the difficulty of measuring the line position on photographic plates.

The first space based CCD spectrum was obtained by \cite{Barstow05} using STIS, and greatly improved the precision of the measurement to 80.42 $\pm$ 4.83 km s$^{-1}$. However, the mass calculated in that analysis could not be reconciled with the spectroscopic mass to confirm consistency between the three methods. A further set of STIS observations were carried out in 2013 \citep{Barstow_2017} and improved the precision further. Unfortunately, the data confirmed there was a definite discrepancy between the gravitational and dynamical mass of around 10 per cent which was larger than the measurement uncertainty.

The question remained as to whether this discrepancy was caused by a fundamental problem with the method, or simply an instrumental effect. Other studies using the gravitational redshift also found a slightly larger average mass in large samples of DA WDs \citep{Falcon_2010} when compared to the average mass of spectroscopic samples. Some of this overestimate may have been due to the contribution of the Stark pressure shift which can produce an additional shift in WD absorption lines (\citealt{Wiese_Kelleher71, Grabowski75}). Laboratory based studies simulating the WD plasma environment have demonstrated that this effect could account for up to 50 per cent of the measured redshift depending on the lines used and the amount of the broad line wings included when measuring the wavelength (e.g. \citealt{Grabowski87, Falcon15_ApJ,Halenka2015}). 

Although the gravitational redshift method has been successfully applied to many WDs, the problem with the Sirius B measurement, which should be the most accurate of all due to its close proximity, continues to cast doubt on this technique. A resolution requires observations which can disentangle any potential systematic effect from the genuine gravitational redshift signal. Once the size and cause of any systematics have been identified, Sirius B can be used to validate the three mass measurement methods and provide an important test of the high mass end of the MRR.

In this paper we present new \textit{HST} observations of Sirius A and B which are used to carry out a differential measurement of the shift in the wavelength of the H$\alpha$ line. These observations have been performed by observing Sirius A and B with the same instrument setup over a single \textit{HST} orbit so that it is possible to distinguish systematic instrumental effects from the gravitational redshift. We make a high precision measurement of the mass of Sirius B which is compared to the dynamical mass and the predictions of the MRR.

\section[]{Observation strategy}

The data for this study were obtained as part of GO program 15237, (PI Joyce) in cycle 25 at the start of 2018. The data consist of 4 exposures each for Sirius A and B. All spectra were taken with the G750M grating which covers the wavelength range 6295 - 6867 \AA\ and captures the broadened wings of the WD H$\alpha$ line centred at $\sim$ 6564 \AA . This set up has a resolution of 0.56 \AA / pixel. This resolution is lower than the approximately 0.26 \AA\ which is the size of the discrepancy from the 2013 observations. However, by fitting a model to the broadened line, cross correlation improves the accuracy of the wavelength measurement by a factor of 10 \citep{Barstow05}. 

Exposure times were calculated to give a S/N $>$ 100 for Sirius B. For Sirius A, the target is bright enough to saturate within 0.1 s. This is shorter than the minimum exposure time of 0.3 s which is limited by the shutter speed. Previous observations \citep{Bohlin_2014AJ_SiriusA} have shown that the spectrum can be recovered even though it is saturated.

A major challenges with this observation is the close proximity of Sirius A which has the potential to contaminate the spectrum of Sirius B. To avoid this problem, the narrow 52$^{\prime\prime}$x0.05 slit was used to exclude light from Sirius A. Also, the orientation of the spacecraft was selected so that the long slit would be perpendicular to the line joining Sirius A and B. This ensures that the slit does not go across both stars and also places Sirius B in between the diffraction spikes caused by the mirror support structure.

The position of the target along the slit changes due to the use of both the standard position, with the spectrum at the centre of the CCD (row 512), and the pseudo E1 aperture position which places the spectrum closer to the top of the CCD (row 898). 

The reason for the use of the E1 position is that it places the spectrum closer to the readout node at the edge of the CCD and minimises loss of signal due to charge transfer inefficiency when the chip is read out \citep{Friedman05_stis_inst_report}. The increasing charge transfer losses as the chip suffers from radiation damage mean that the E1 pseudo aperture is now the preferred position, except for bright targets which are likely to saturate the detector. 

The Sirius A spectrum was likely to be highly saturated. For such a saturated spectrum, it is recommended to use the original (centre) position \citep{Friedman05_stis_inst_report}.  
It was decided to take some exposures of each target at both slit positions (E1 and centre) so that any effects due to charge transfer inefficiency losses and saturation could be compared since it was not known how these might affect the absolute wavelength calibration of the spectrum. 

The sequence of exposures is listed in Table \ref{table:observations_2018}.
When using the narrow (52$^{\prime\prime}$x0.05) slit, normal procedure is to perform a peak-up to precisely centre the target in the slit. This was done for Sirius B at the start of the observing run. The telescope was then moved to point at Sirius A. However, Sirius A is too bright to perform a peak up without changing the slit and filter settings. The distance moved between targets is only 11 arc seconds and the precision of the telescope pointing for such a small angle manoeuvre \footnote{A 3 arcsecond manoeuvre has an error of 0.003 arcsecond according to the STIS Instrument Handbook \citep{Riley_2017}, Section 8.2.3} is 0$^{\prime\prime}$.0045  It was therefore decided to forgo the peak up procedure so as to take an exposure of Sirius A without any intervening change to the slit or filter. The first sequence of spectra (odl601010 to odl601040) are therefore identical in terms of instrument set up and almost co-incident in time. 

The second set of Sirius A exposures (odl601050 / 060) was taken after a peak-up using the ND filter to ensure the target was correctly aligned in the slit and placed at the centre position on the CCD. The following Sirius B exposure (odl601070) was also taken at the centre position for direct comparison with the preceding Sirius A spectrum (odl601060). Finally, one more spectrum of Sirius B was taken back at the E1 position. This can be compared to the two Sirius B exposures taken at the start of the orbit to check for any shift in the H$\alpha$ line over the course of the observing run due to instrumental changes or possibly thermal effects such as heating and cooling of the optical bench. 

\begin{figure}
\includegraphics[width=\linewidth]{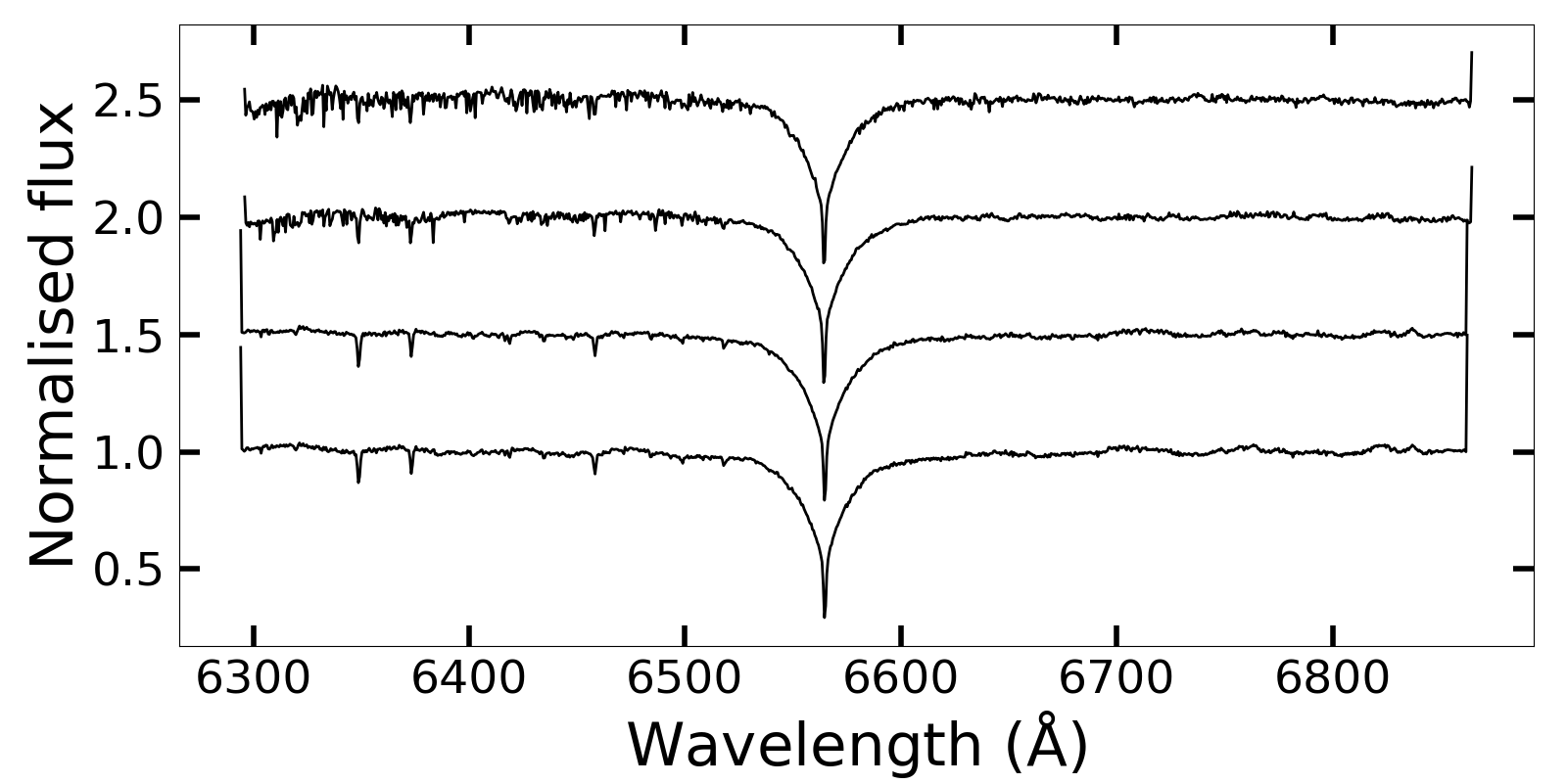}
  \caption{Sirius A H$\alpha$ line. The 4 spectra have been normalised to a flat continuum and are offset by 0.5 in flux for clarity. }
  \label{fig:Sirius_A_normalised_full}
\end{figure}

\begin{figure}
\includegraphics[width=\linewidth]{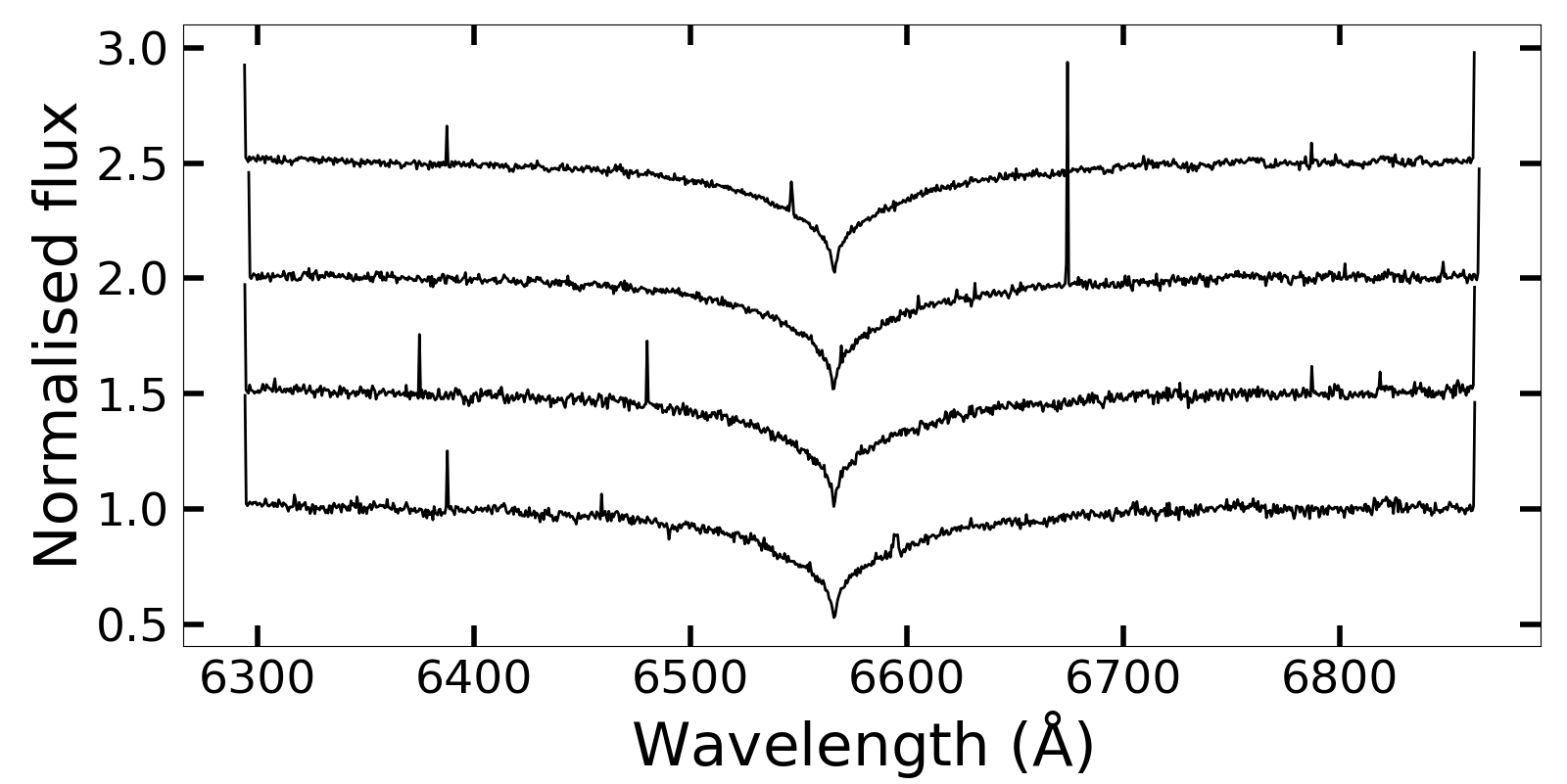}
  \caption{Sirius B H$\alpha$ line. The 4 spectra have been normalised to a flat continuum and are offset by 0.5 in flux for clarity. }
  \label{fig:Sirius_B_normalised_full}
\end{figure}

The four spectra for each target are plotted in Fig. \ref{fig:Sirius_A_normalised_full} and \ref{fig:Sirius_B_normalised_full}. The spectra have been normalised to remove the slope of the continuum and are offset by increments of 0.5 on the y axis for clarity. The core region of the H$\alpha$ line in all spectra used for the measurements are free of cosmic ray hits and show no signs of any peculiarities which might affect the wavelength measurements.

\begin{table}

\caption{Spectra of Sirius A and B taken with \textit{HST} on 12/01/2018. Program 15237, (PI Joyce, Barstow) }
\label{table:observations_2018}
\begin{tabular}{cccc}

\hline
 Obs ID & CCD position & Exp time  & Time of exp  \\
& (Centre or E1) & (s) & (GMT)\\
\hline
\textbf{Sirius B} &  &  &  \\

ODL601010 & E1 & 100 & 03:32:40\\
ODL601020 & E1 & 100 & 03:36:14\\

\hline
\textbf{Sirius A} &  &  & \\

ODL601030 & E1 & 0.2 & 03:39:59 \\
ODL601040 & E1 & 0.2 & 03:41:29\\
ODL601050 & Centre & 0.9 & 03:53:36\\
ODL601060 & Centre & 0.9 & 03:56:17\\

\hline

\textbf{Sirius B} &  &  &  \\
ODL601070 & Centre & 75 & 04:01:35 \\
ODL601080 & E1 & 75 & 04:04:16\\

\hline

\end{tabular}
\end{table}


\section{Analysis}

\subsection{Fitting method and corrections}

\subsubsection{Check for contamination}

\begin{figure}
\includegraphics[width=1\linewidth]{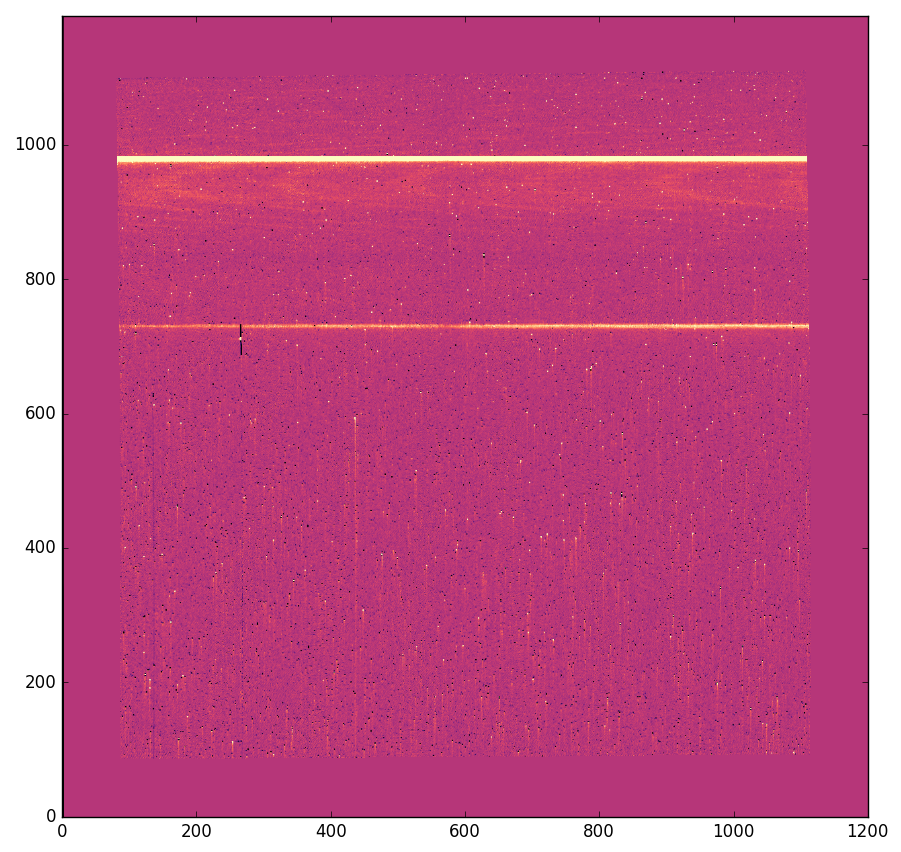} 

\includegraphics[width=1\linewidth]{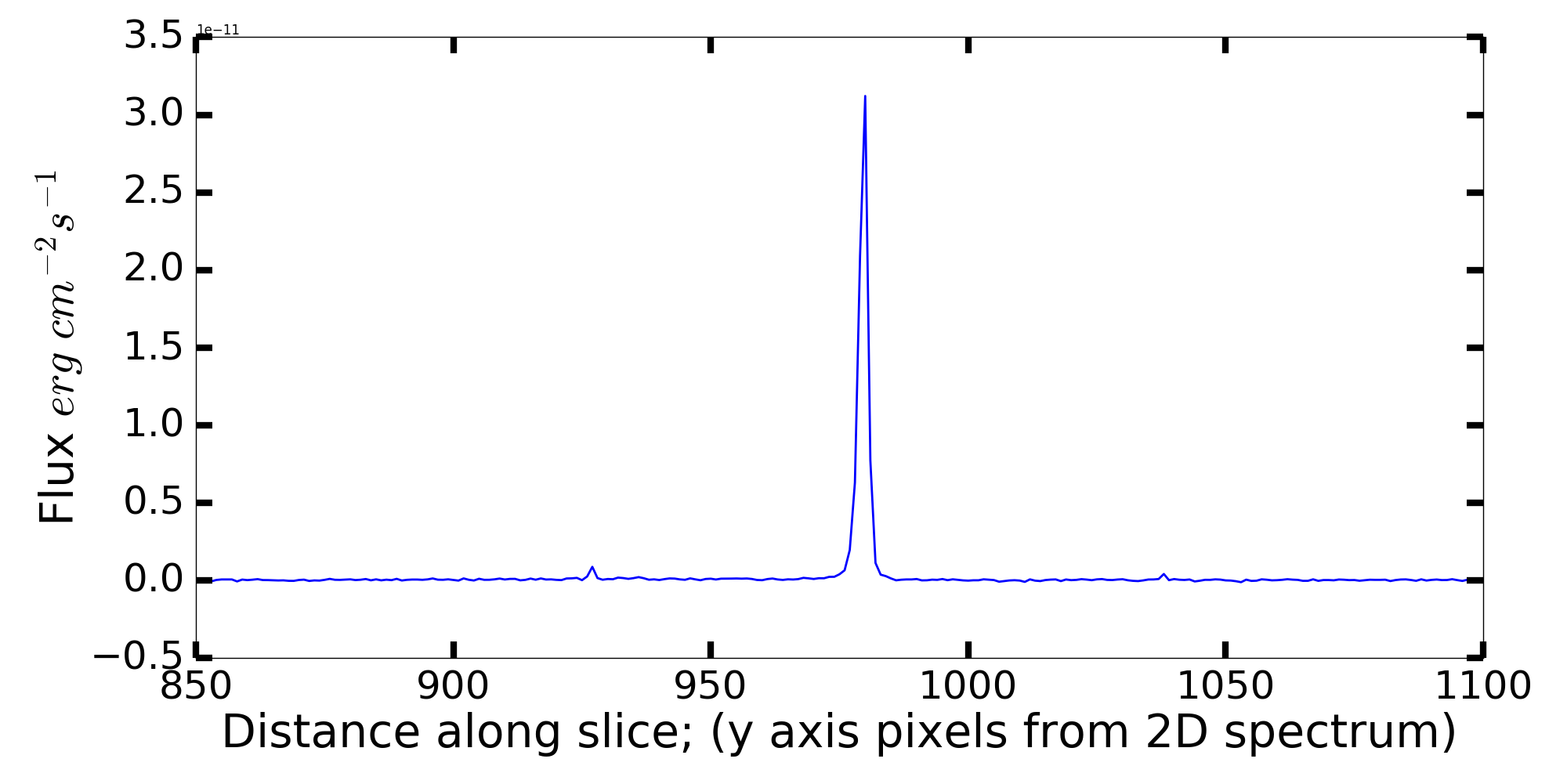}

\caption{Examples of the 2-D spectra of Sirius B showing the position of the spectrum on the CCD at the E1 position. The images are histogram equalised to show up any background clearly. Faint lines either side of the main spectrum are the signal from the diffraction spikes of Sirius A. The lower panel shows the flux for a 1 pixel wide vertical slice through the 2-D spectrum.}\label{fig:Sirius_B_2d_spec_E1}
\end{figure}

\begin{figure}

\includegraphics[width=1\linewidth]{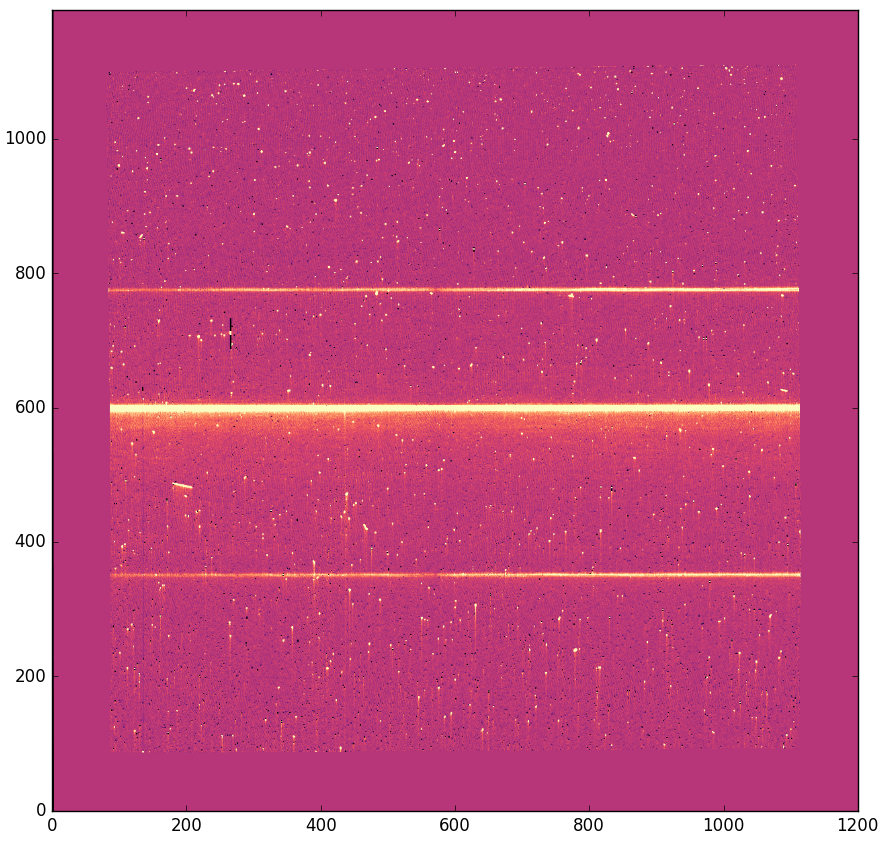} 
\includegraphics[width=1\linewidth]{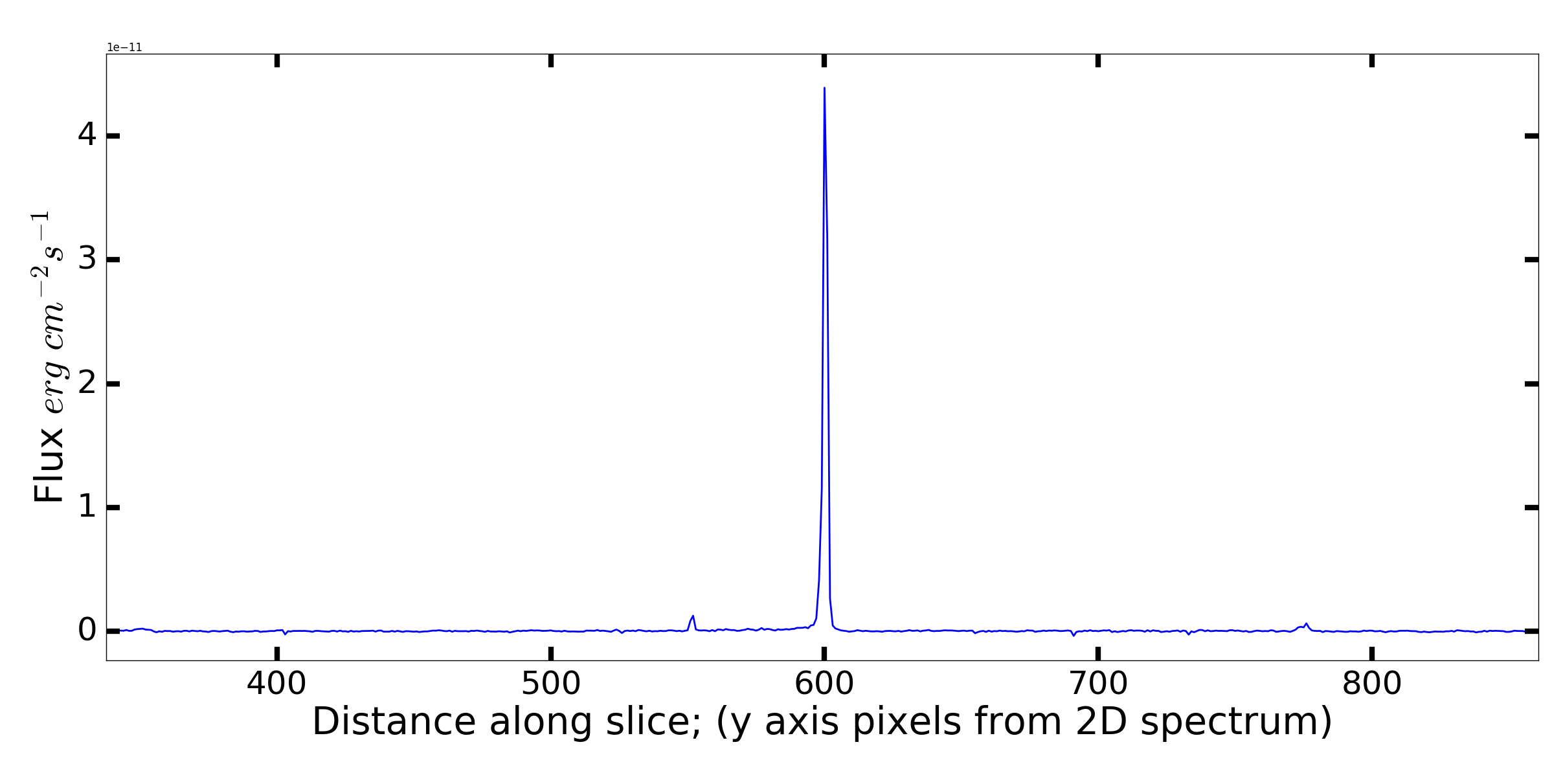} 

\caption{Same as Fig. \ref{fig:Sirius_B_2d_spec_E1} but for a spectrum taken at the centre of the CCD rather than the E1 position.}\label{fig:Sirius_B_2d_spec_centre}
\end{figure}

The spectra of Sirius B were checked for any signs of contamination by light from Sirius A. In Fig. \ref{fig:Sirius_B_2d_spec_E1} and \ref{fig:Sirius_B_2d_spec_centre} are examples of the 2D images from which the Sirius B spectra are extracted. They show no significant signs of scattered light from Sirius A except for the faint spectra either side of the bright Sirius B spectrum due to the diffraction spikes crossing the slit. The lower panel in each figure shows a vertical slice through the 2-D image with the spike in the flux at the position of the Sirius B spectrum. This shows that the spectrum is unaffected by scattered light and the faint spectrum from the diffraction spikes has no impact on the main spectrum. Fig. \ref{fig:Sirius_B_2d_spec_E1} and \ref{fig:Sirius_B_2d_spec_centre} also illustrate the difference in position between spectra taken with the E1 and centre settings.


\subsubsection{Correction for \textit{HST} orbital motion}

\begin{figure}
\includegraphics[width=\linewidth]{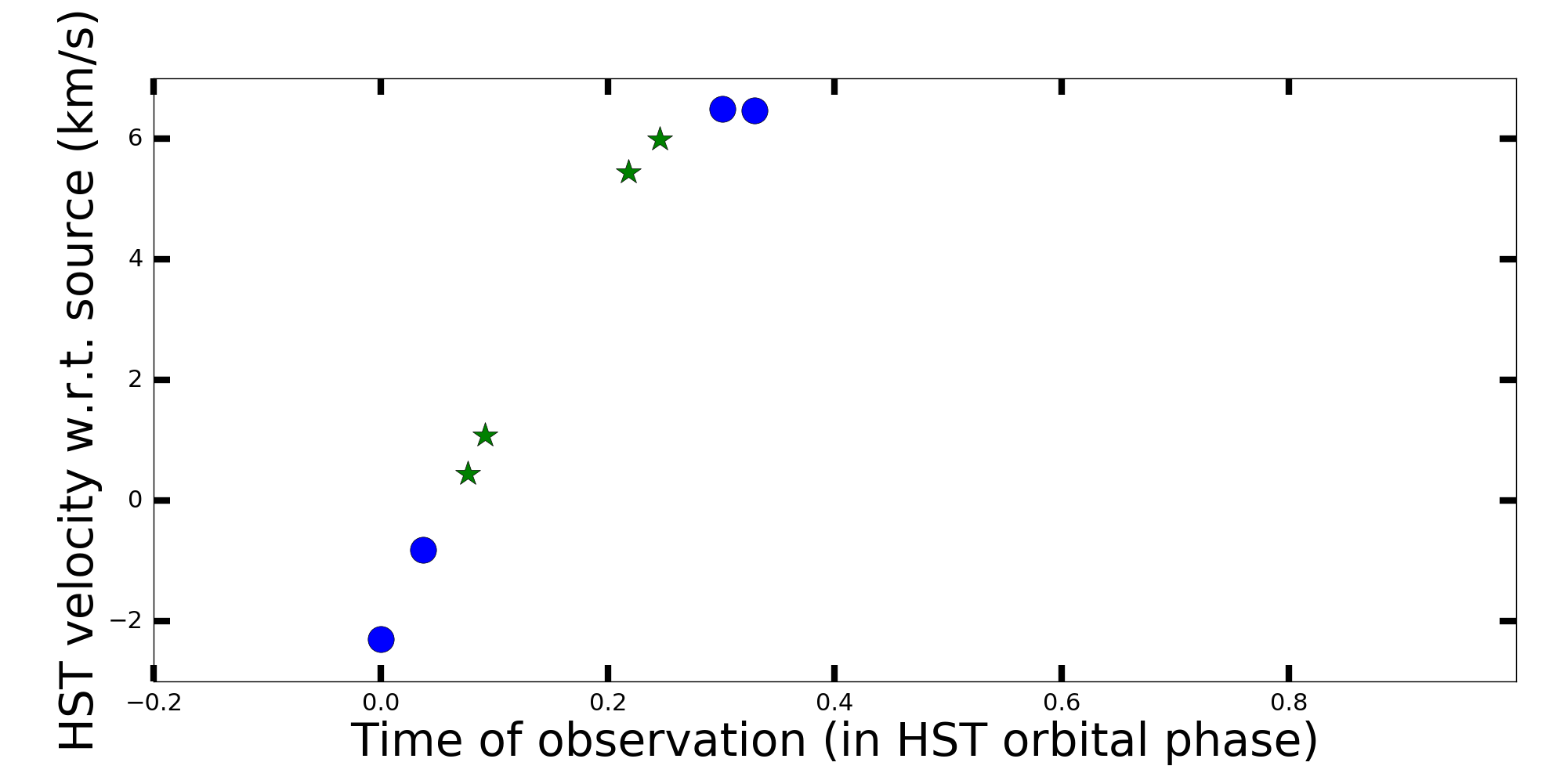}
  \caption{Orbital velocity of the \textit{HST} relative to the target during each of the exposures. Symbols indicate the target : Sirius B (circles), Sirius A (stars)}
  \label{fig:HST_orbital_velocity_2018}
\end{figure}

An additional shift is added to the observed spectra by the orbital motion of the \textit{HST} with respect to the target. This correction is not applied in the pipeline processing and can alter the measured velocity by up to $\pm$7.5 km s$^{-1}$. The magnitude of the effect varies over the course of the orbit so a different correction must be applied to each exposure individually.
The orbital velocity at the time of each exposure is plotted in Fig. \ref{fig:HST_orbital_velocity_2018} and values are listed in Table \ref{table:correction_factors_2018} (column 2). The corresponding shift in the wavelength was calculated using equation (\ref{eq:velocity_from_wave}) which converts the \textit{HST} velocity to a corresponding shift between points A and B on the detector. The correction is applied directly to the spectral files before model fitting.

\begin{equation}\label{eq:velocity_from_wave}
v_{\rm obs} = \frac{\lambda_{B} - \lambda_{A}}{\lambda_{A}} \times c
\end{equation}

\subsubsection{Measuring the wavelength of H$\alpha$}

The wavelength of the observed H$\alpha$ line in the Sirius A and B spectra was measured by fitting a Lorentzian model to the core of the line. A python script utilizing the 'lmfit' library \citep{lmfit} was used to perform the fitting. The 'lmfit' fitting function uses the least-squares method of Levenberg-Marquardt \citep{Press86}. This approach differs from the method used previously \citep{Barstow05, Barstow_2017} because it does not rely on fitting with \textsc{tlusty} models which have an inbuilt rest wavelength for H$\alpha$ and only provide the shift relative to this standard of rest. Here, the fitting makes no assumption about the rest wavelength and is simply a measure of the wavelength at the centre of the line.

Fig. \ref{fig:Sirius_A_normalised_full} and \ref{fig:Sirius_B_normalised_full} show that the H$\alpha$ line of Sirius B is broader than Sirius A and covers a range of $\sim300$ \AA. To accurately measure the wavelength of the line centre, the fitting for each line was repeated four times with a slightly increased wavelength range each time. The ranges used are 7,11,15 and 19 \AA.  These were chosen so as to focus on the sharply defined line core and avoid including too much of the wings which may be affected by the Stark pressure shift and asymmetry. Tests of lab based plasma have shown that the Stark shift in the H$\alpha$ line increases with increasing distance from the line core \citep{Halenka2015}. For the wavelength ranges we have chosen (7-19 \AA), the effect of the Stark shift is expected to be below 1 km s$^{-1}$ based on simulations and observations of plasma in the laboratory (\citealt{Halenka2015}, see their fig. 13).


\begin{figure}
\includegraphics[width=1\linewidth]{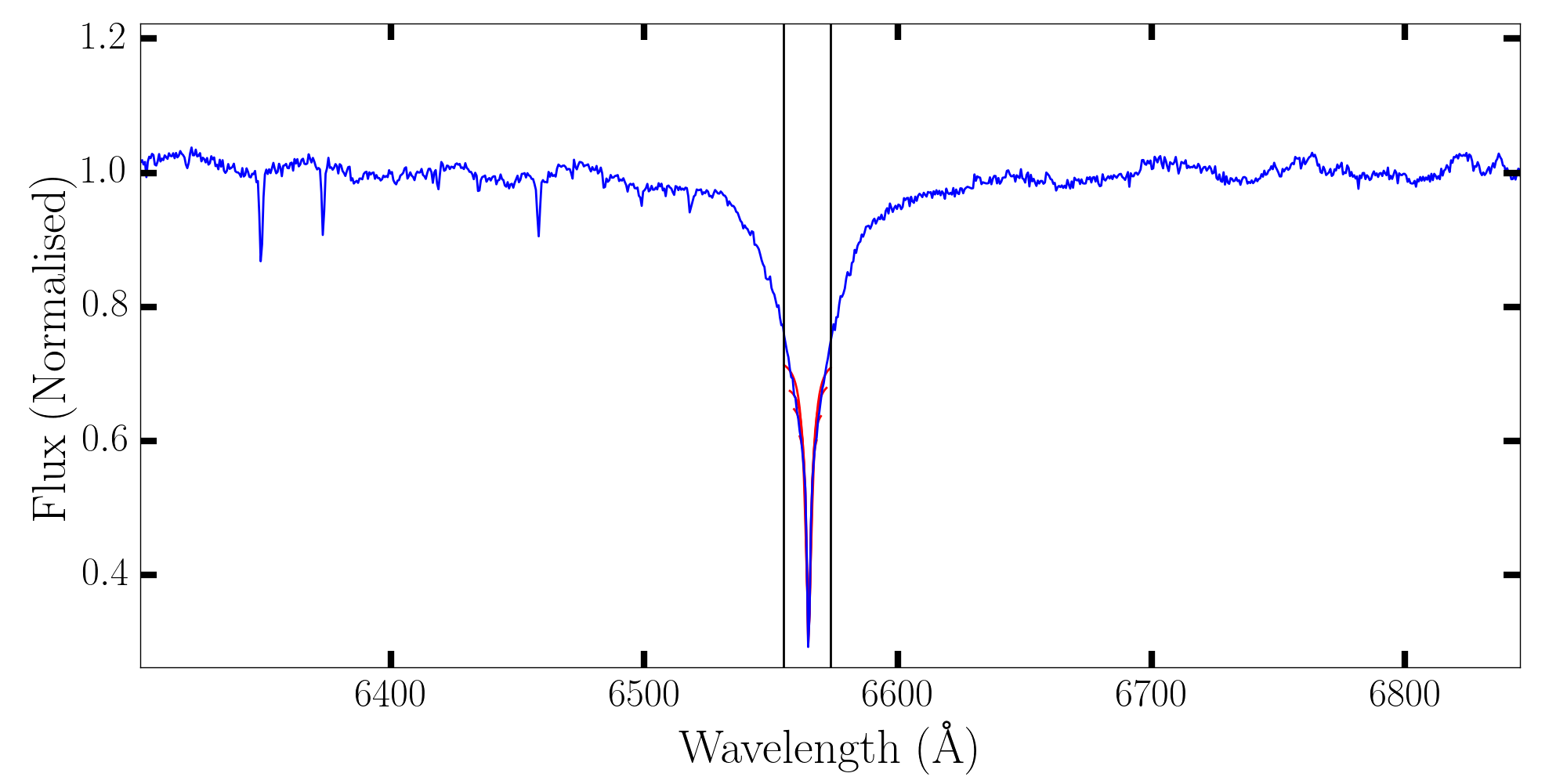} 
\includegraphics[width=1\linewidth]{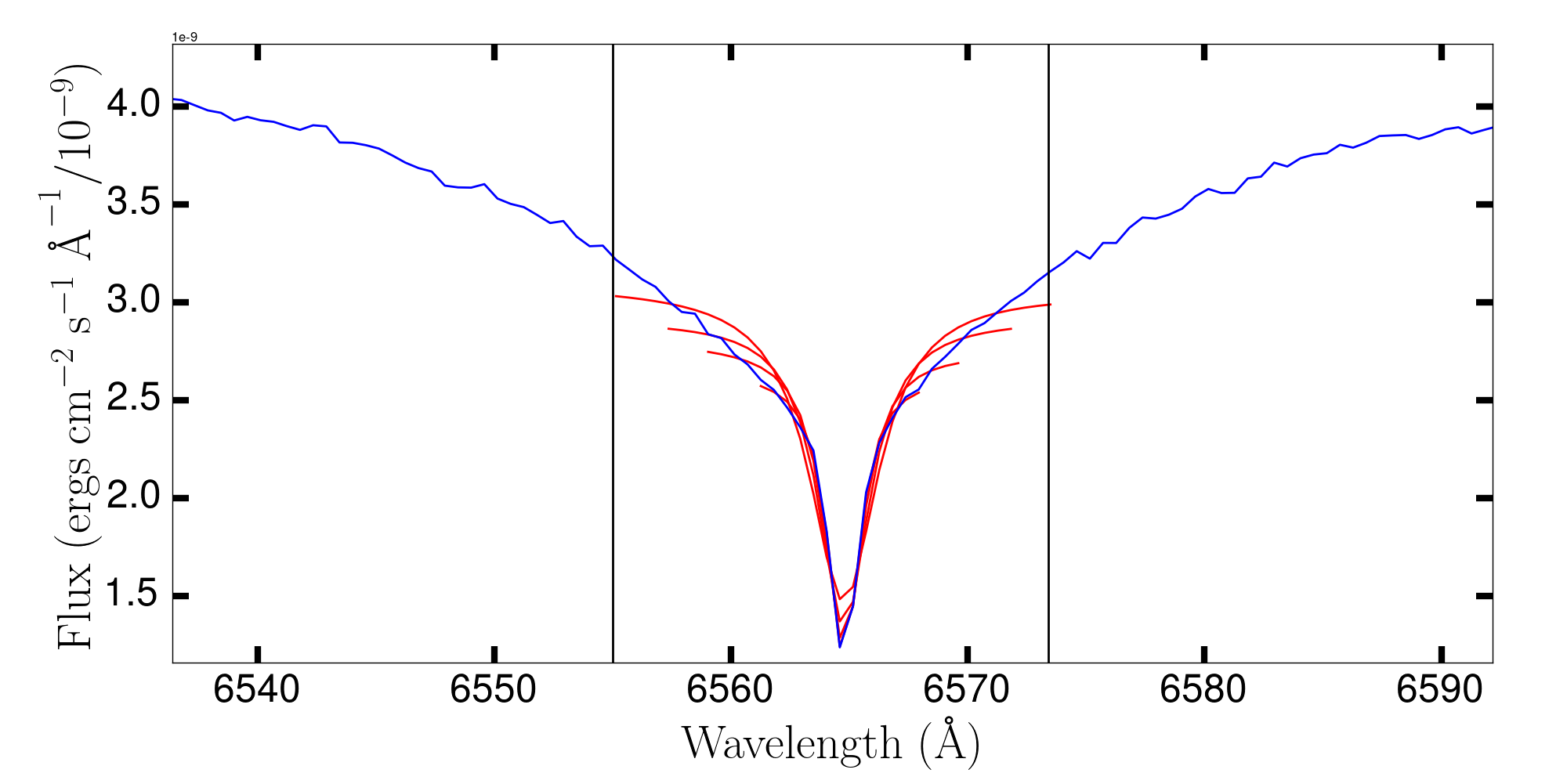} 

\caption{\textbf{Top panel:} Sirius A, showing the full H$\alpha$ line with the core of the line used for fitting indicated by the two horizontal lines.
\textbf{Bottom panel:} Sirius A, zoom in on the central region of the figure above to show the detail of the fitting using the four wavelength ranges.}
\label{fig:Sirius_A_fitting}
\end{figure}



\begin{figure}
\includegraphics[width=1\linewidth]{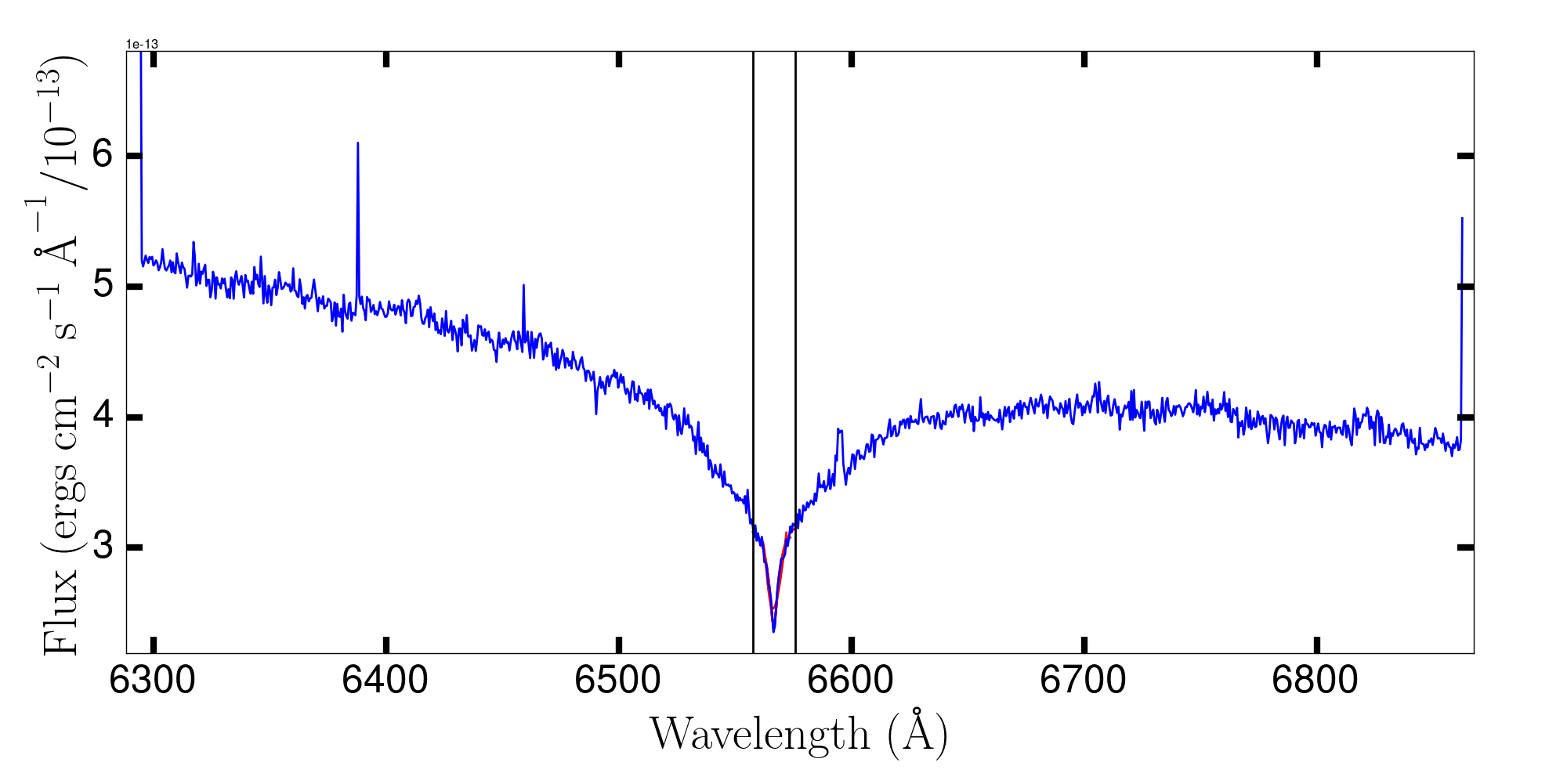}
\includegraphics[width=1\linewidth]{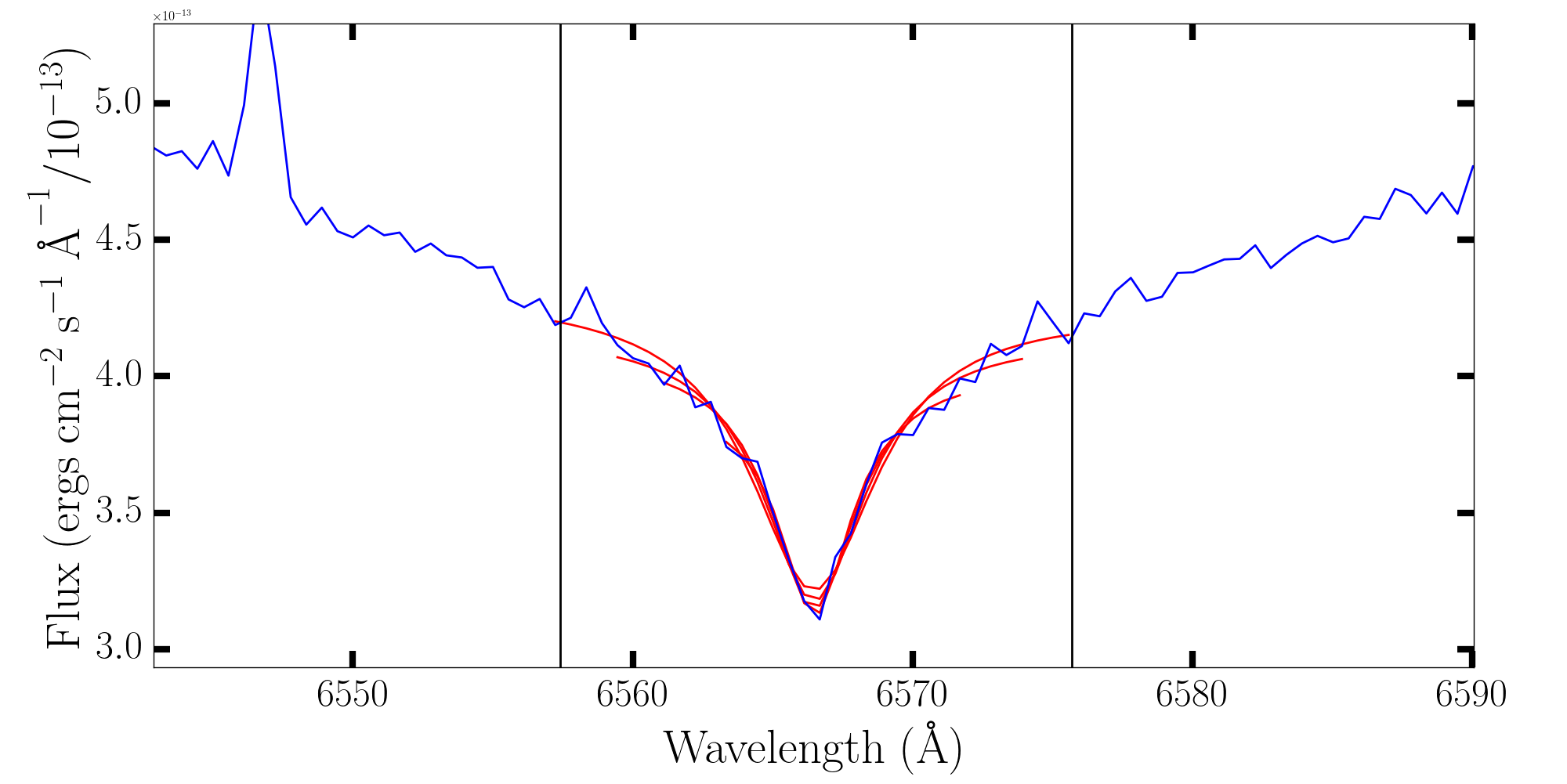} \includegraphics[width=1\linewidth]{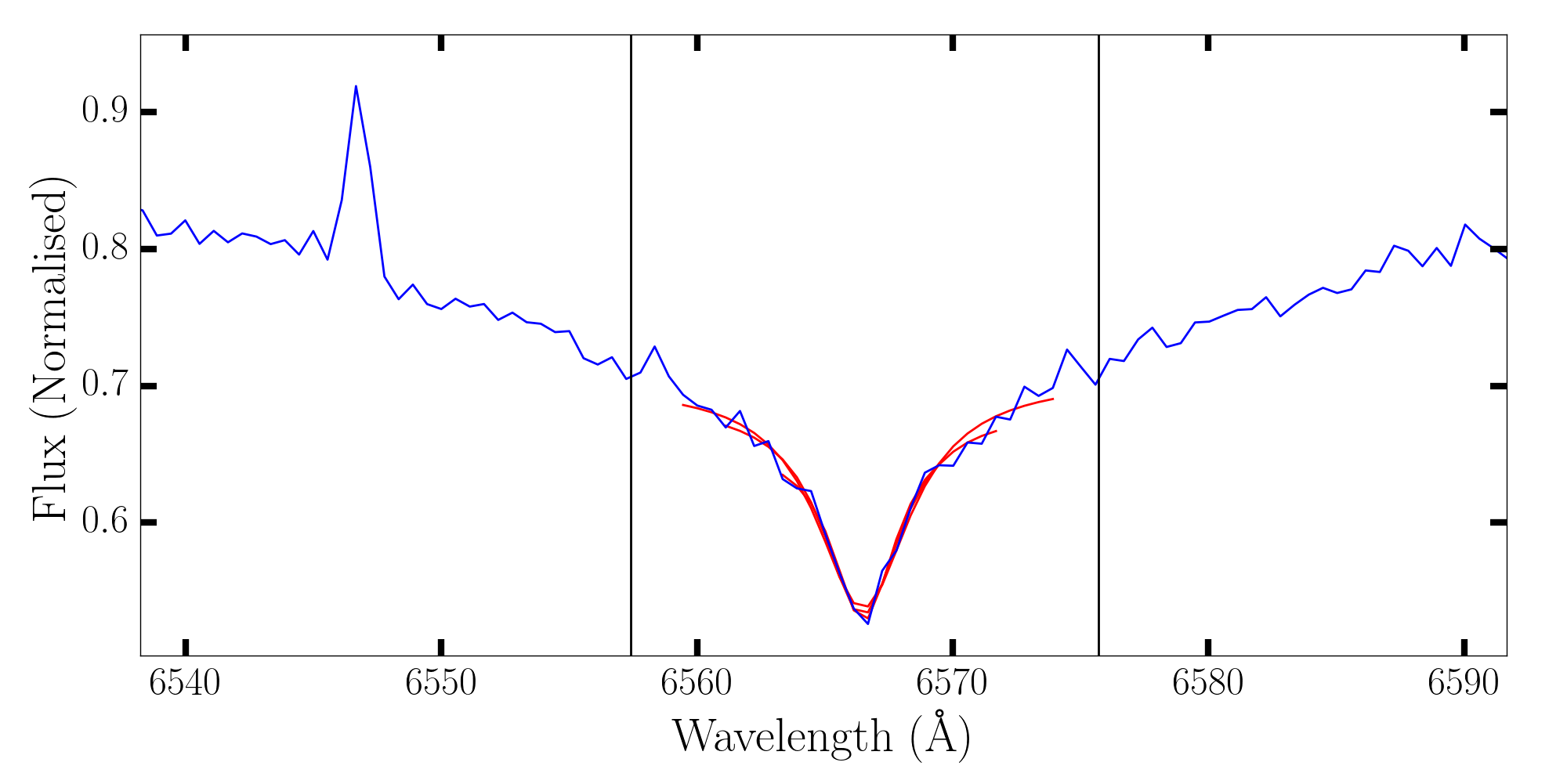} 
	
\caption{Example of fitting the Sirius B H$\alpha$ line to measure the wavelength. \textbf{Top panel:} The full wavelength range and the extent of the broad wings. \textbf{Middle panel:} A zoom in to show detail of the fitting using the four wavelength ranges which only include the core of the line. \textbf{Bottom panel:} Same as the middle but the spectrum has been normalised to remove the slope of the continuum.}
\label{fig:Sirius_B_fitting}
\end{figure}


\subsubsection{Calculation of the velocity}

\begin{figure}
\includegraphics[width=\linewidth]{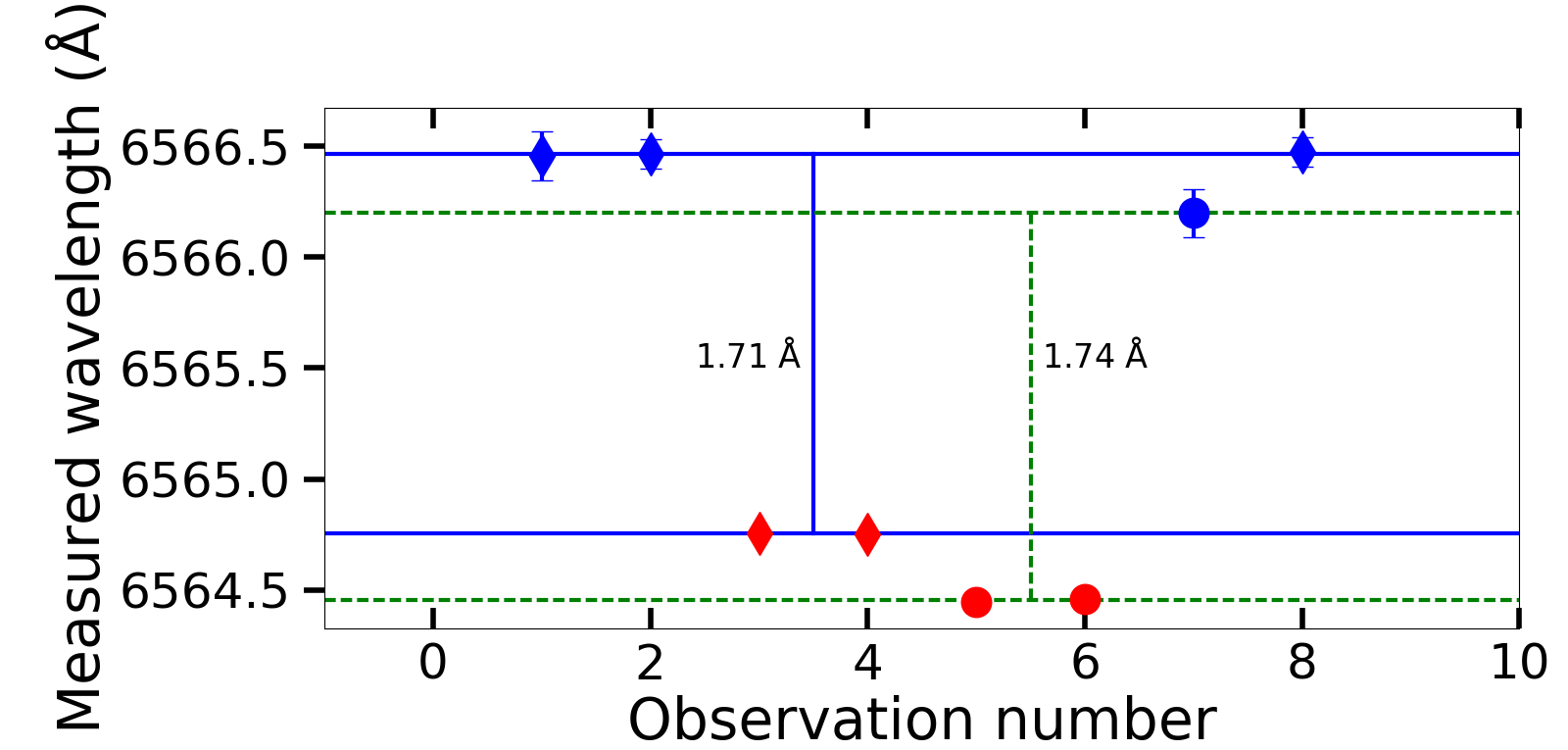}
  \caption{Measured wavelength for each spectrum in the order in which they were observed along the x axis. Colours indicate the target. Sirius A (red, lower markers) and Sirius B (blue, upper markers). Shapes indicate the aperture used, E1 position (diamonds) and 'centre' (circles). Horizontal lines show the average wavelength and vertical lines indicate the difference in average wavelength between Sirius A and B.}
  \label{fig:raw_wavelength}
\end{figure}

The observed velocity is calculated from the difference in the measured wavelengths between Sirius A and Sirius B. Fig. \ref{fig:raw_wavelength} shows the measured wavelength of each H$\alpha$ line in the order in which they were observed. The markers are blue for Sirius B and red for Sirius A. This clearly highlights the difference caused by using the E1 position (diamonds) compared to the centre position (circles). The horizontal lines indicate the average wavelength measured for the E1 spectra (blue, solid line) and centre spectra (green, dashed line). The wavelengths from all individual spectra are consistent with the average when sorted by target and CCD position. The wavelength difference between A and B is calculated using the average wavelengths. E1 and centre data were calculated separately. The horizontal lines show that the difference in wavelength between A and B is 1.71 \AA\ for E1 and 1.74 \AA\ for centre. The wavelength difference is converted into a velocity using equation (\ref{eq:velocity_from_wave}).

\subsubsection{Correction for velocity of Sirius A and B}

\begin{figure}
\includegraphics[width=\linewidth]{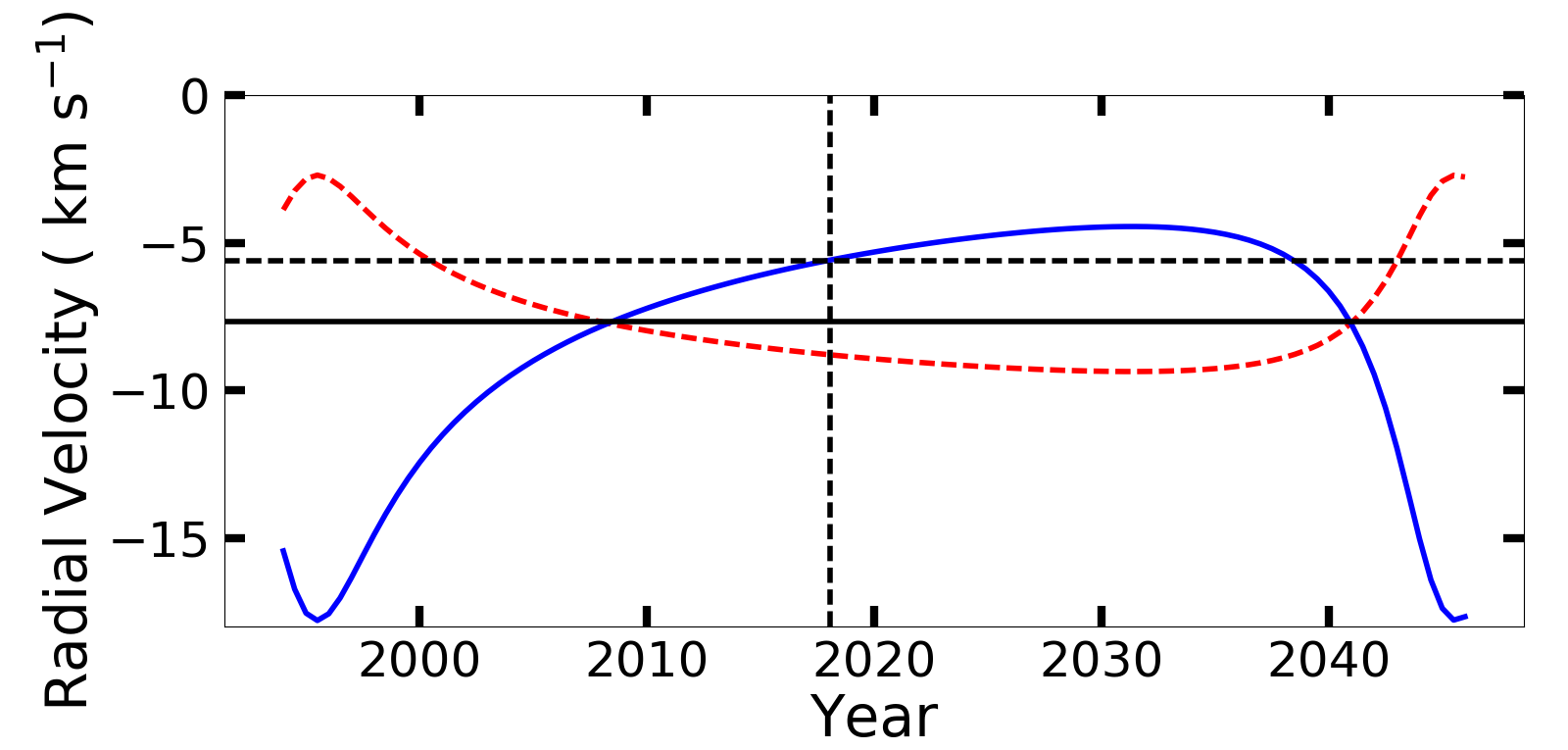}
  \caption{Velocity of Sirius A and B at the time of the 2018 observations. The dashed red curve (Sirius A) and blue curve (Sirius B) show the velocity of each star relative to the observer. The dashed black line marks the Sirius B velocity during the 2018 observations. The solid horizontal line is the velocity of the binary centre of mass ($\gamma$).}
  \label{fig:Sirius_RV_2018}
\end{figure}

Fig. \ref{fig:Sirius_RV_2018} shows the radial velocities for Sirius A and B extrapolated from the astrometrically determined orbit \citep{Bond_sirius_17}. The $\gamma$ velocity of the binary centre of mass is -7.687 km s$^{-1}$ marked by the solid black line. At the time of the 2018 observations, the velocity relative to the observer was -5.596 km s$^{-1}$ for Sirius B and -8.794 km s$^{-1}$ for Sirius A. These values include the $\gamma$ velocity. 

The $\gamma$ velocity affects both stars equally so it does not affect the relative wavelength shift. The only velocity that needs to be considered is the difference due to the orbital motion of the 2 stars ($K_{\rm diff}$ velocity). The difference in velocity between A and B is $-8.794 - (-5.596) = -3.198$ km s$^{-1}$ i.e., a net differential velocity towards the observer. This has the effect of reducing the observed wavelength shift (a blue shift). The velocity of 3.198 km s$^{-1}$ must be added back on to the observed velocity. 

In addition to this, there is a small correction for the gravitational redshift of Sirius A. Sirius A is large and close enough to have had its angular diameter measured directly by interferometry at 5.936 $\pm 0.016$ mas \citep{Kervella_2003}. At the distance of 2.36 pc this gives a radius of R = 1.711 $\pm$ 0.013 R$_{\odot}$. From equation \ref{eq:mass_from_Vgr}, using a dynamical mass of 2.042 $\pm$ 0.01 M$_{\odot}$ \citep{Bond_sirius_17}, this gives a gravitational redshift of 0.759 km s$^{-1}$ for Sirius A. This produces an additional redshift velocity which must be subtracted. The final velocity is calculated using equation (\ref{eq:corrections_to_differential_velocity}).

\begin{equation}\label{eq:corrections_to_differential_velocity}
v_{\rm gr} = v_{\rm obs} + K_{\rm diff} - V_{\rm MS,gr}
\end{equation}

\begin{table}

\caption{List of correction factors calculated for Sirius A and B.}
\begin{tabular}{ccc}

\hline

Obs ID & HST Orbital velocity & slit angle offset\\
 &  (km s$^{-1}$) & (km s$^{-1}$) \\
\hline 
\textbf{Sirius A} &    &  \\ 

odl601030 & 0.4368 & 0.0  \\ 

odl601040 &  1.076 & 0.0  \\ 

odl601050 & 5.441 & -0.634 \\ 

odl601060  & 5.985 & 0.634  \\ 
\hline 

 \textbf{Sirius B} &  &  \\ 

odl601010 & -2.299 & -0.63  \\ 

odl601020  & -0.816 & 0.63 \\ 

odl601070  & 6.497 & 0.0 \\ 

odl601080  & 6.468 & 0.0 \\ 

\hline 
\end{tabular}\label{table:correction_factors_2018}
\end{table}


\subsubsection{Slit angle correction}

\begin{figure}
\includegraphics[width=0.8\linewidth]{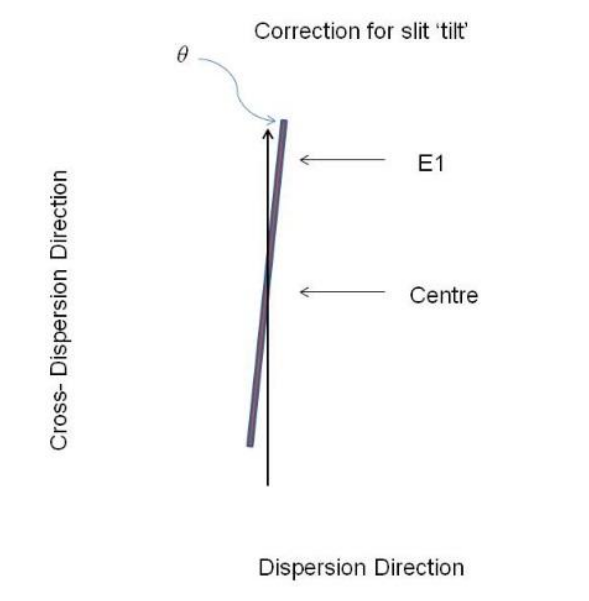}
  \caption{The angle of the 52x0.05" slit causes a slight offset along the dispersion (wavelength) direction when the target is at the E1 position compared to the centre position.}
  \label{fig:schematic_of_slit_angle}
\end{figure}

The 52$^{\prime\prime}$x0.05 slit is not exactly perpendicular to the dispersion direction of the CCD. It is at an angle of 0.35$^{\circ}$ to the perpendicular\footnote{STIS instrument handbook, table 11.2}. The position of the spectrum on the CCD therefore shifts slightly in the dispersion direction depending on the position of the target in the slit as shown in Fig. \ref{fig:schematic_of_slit_angle}. This is important because the target is dithered along the slit between exposures so as to minimise potential problems with hot pixels. The position along the slit is recorded in the file header as \textsc{postarg2} and can be used to calculate the shift in the wavelength. For the dithering pattern specified for these observations, only spectra 1,2,5 and 6 were offset along the slit. 

As an example, for spectrum odl601020 the offset along the slit was set to $\pm$ 0.203120 arcsec. With a plate scale of 0.05 arsec/pixel this is $\pm$ 4 pixels. The offset in the dispersion direction is then $ 4 \times 0.35 \times \frac{\pi}{180} = 0.0248$ (pixels). In the dispersion direction the scale is 0.56 \AA\ per pixel, amounting to a wavelength offset of $0.56 \times 0.0248 = 0.0139$ \AA\ or 0.63 km s$^{-1}$. Values for the offset due to dithering are listed in Table \ref{table:correction_factors_2018}. The wavelength offsets due to dithering are not automatically corrected in the pipeline so they have been included in the corrections to the velocity for each spectrum.

\subsubsection{Mass calculations}

Table \ref{table:results_2018} gives the difference in wavelength between Sirius A and Sirius B as plotted in Fig. \ref{fig:raw_wavelength}. There are separate rows for observations taken at the E1 position and the centre position. The $v_{\rm obs}$ column gives the velocity measured from the difference in wavelength between Sirius A and B before any correction for the motion of the source. The $v_{\rm gr}$ column is the velocity after correcting for relative velocity between A and B and the gravitational redshift of Sirius A. Having removed all additional sources of Doppler velocity, what remains is attributed entirely to the gravitational redshift effect. The final column is the mass calculated from $v_{\rm gr}$ and the radius using equation (\ref{eq:mass_from_Vgr}). In this equation $v_{\rm gr}$ is in km s$^{-1}$ and $M$ and $R$ are in solar units. The radius of 0.803 $\pm$ 0.011 R$_{\odot}$/100 was calculated from the flux in the G430L spectra as described in \cite{Joyce_2018_a} and uses a parallax of 378.9 $\pm$ 1.4 miliarcseconds from \cite{Bond_sirius_17}.

\begin{equation}\label{eq:mass_from_Vgr}
M = \frac{v_{gr} R}{0.636}   
\end{equation}

\begin{table*}

\caption{Measured wavelength, velocity and mass.}
\begin{tabular}{cccccc}

\hline

obsID & wavelength & $\Delta$ $\lambda$ & $v_{obs}$ & $v_{gr}$ & Mass\\
 & ( \AA\ ) & ( \AA\ ) & (km s$^{-1}$)  &  (km s$^{-1}$) & (M$_{\odot}$)\\
\hline 
\textbf{E1} &  &  & && \\
\hline 
\textbf{Sirius A} &  &  & & &  \\ 
odl601030 & 6564.756 $\pm$ 0.010 & - & - & - & - \\ 
odl601040 & 6564.750 $\pm$ 0.015 & - & - & - & - \\ 
Average & 6564.753 $\pm$ 0.002 &&&&\\
&&&&&\\
\textbf{Sirius B} &  &  & & &  \\ 

odl601010 & 6566.456 $\pm$ 0.111 & - &-& - & - \\ 

odl601020 & 6566.466 $\pm$ 0.068 & - &-& - & - \\ 

odl601080 & 6566.476 $\pm$ 0.067 & - &-& - & - \\ 
Average & 6566.466 $\pm$ 0.006 &&&&\\
&&&&&\\
\hline
Differential & & 1.713 $\pm$ 0.006 & 78.21 $\pm$ 0.28 & 80.65 $\pm$ 0.77 & 1.017 $\pm$ 0.025 \\
\hline
&&&&&\\
&&&&&\\

\hline 
 \textbf{Centre} &  &  &  & & \\ 
\hline  
  \textbf{Sirius A} &  &  & &  & \\ 

odl601050 & 6564.447 $\pm$ 0.015 & - &-& - & - \\ 

odl601060 & 6564.459 $\pm$ 0.011 & -& -& - & - \\ 
Average & 6564.453 $\pm$ 0.004 &&&&\\
&&&&&\\
\textbf{Sirius B} &  &  & &  & \\ 
odl601070 & 6566.199 $\pm$ 0.108 & -&- & - & - \\ 
&&&&&\\
\hline 
Differential && 1.746 $\pm$ 0.004 & 79.73 $\pm$ 0.19 & 82.17 $\pm$ 0.74 & 1.036 $\pm$ 0.025 \\

\hline 
\end{tabular}\label{table:results_2018}
\end{table*}

The wavelength measured for the H$\alpha$ line in each spectrum is listed in Table \ref{table:results_2018}. From the average of the E1 spectra, the observed wavelength is 6564.753 $\pm$ 0.002 \AA\ for Sirius A and 6566.466 $\pm$ 0.006 \AA\ for Sirius B. The difference between the measured wavelengths is 1.713 $\pm$ 0.006 \AA\ which gives an observed velocity of 78.31 $\pm$ 0.28 km s$^{-1}$. 

A correction of +2.44 km s$^{-1}$ was applied to remove the effect of the velocity of the source and the gravitational redshift of Sirius A. The final velocity of 80.65 $\pm$ 0.77 km s$^{-1}$ gives a mass of 1.017 $\pm$ 0.025 M$_{\odot}$ via equation (\ref{eq:mass_from_Vgr}). For the spectra taken at the centre of the CCD, the same process gives a slightly larger mass of 1.036 $\pm$ 0.025 M$_{\odot}$.


\subsection{Comparison of fitting methods}

The mass of Sirius B calculated via the gravitational redshift is very sensitive to the measured wavelengths of the H$\alpha$ lines. We have carried out tests using 3 slightly different methods of measuring the wavelength to ensure that the results are robust and check for any bias introduced by the fitting method.

The first method is to fit Lorentzian profiles to the core of the line as previously described. This method was applied in exactly the same way to Sirius A and B to ensure consistent results. One possible source of bias is that the H$\alpha$ line is affected by the slope of the continuum and is therefore not exactly symmetrical. To check if this has any significant effect on the wavelength, the spectra were all normalised to remove the slope of the continuum. The flattened spectra were then fitted in the exact same way as the non-normalised data and the measured wavelengths compared (see Table \ref{table:wave_method_compare}, Method 2).

The third method of fitting is the same as that applied to the 2013 data in \cite{Barstow05, Barstow_2017}, which makes use of a white dwarf spectral model calculated for the $T_{\rm eff}$ and log $g$ of Sirius B as found from spectroscopic fitting of the Balmer lines.
This method differs from the previous two in that it uses a model specifically calculated to match the line broadening in a white dwarf atmosphere and is therefore able to simultaneously fit both the core and wings of the line more accurately than a Lorentzian model. When fitting with this model, broader wavelength regions are included so as to include more of the wings. The ranges are 7,64,120 and 176 \AA\  The disadvantage of this method is that it does not provide the wavelength of the line directly. The model includes a $z$ parameter which measures the shift required to fit the data compared to the model standard of rest. The model uses a rest wavelength of 6562.795 \AA\ which is appropriate for measurements in air \citep{NIST}. However \textit{HST} operates in a vacuum so the rest wavelength should be 6564.6078 \AA\ . The measured wavelengths are adjusted to correct for this difference. The models are also sensitive to the values chosen for the $T_{\rm eff}$ and log $g$ parameters which are fixed at the values found from fitting the G430L data. It has already been shown that the mass calculated from the spectroscopic method is lower than expected which may mean that improvements to the models are required which could result in different best fitting $T_{\rm eff}$ and log $g$ values.

\begin{table*}

\caption{Comparison of the wavelength measured for the Sirius B H$\alpha$ line as measured using 3 different methods.}
\begin{tabular}{ccccc}

\hline

obsID & Method 1 & Method 2 & Method 3 & Standard\\
& With cont. & Normalised & \textsc{xspec}&deviation\\
 & (\AA) & (\AA) & (\AA)  &  ($\sigma$) \\

\hline 
odl601010 (E1) & 6566.46$\pm$0.1 & 6566.46$\pm$0.1 & 6566.46$\pm$0.03 & 0.0018 \\
odl601020 (E1) & 6566.47$\pm$0.07 & 6566.48$\pm$0.04 & 6566.43$\pm$0.04 & 0.0198 \\ 
odl601080 (E1) &  6566.48$\pm$0.07  & 6566.48$\pm$0.06 & 6566.49$\pm$0.08 & 0.0050 \\ 
Average & 6566.466 & 6566.474 & 6566.458 &\\

&&&&\\
odl601070 (Centre) &  6566.20$\pm$0.1  & 6566.21$\pm$0.1 & 6566.21$\pm$0.04 & 0.0046 \\ 

&&&&\\

\hline 
\end{tabular}\label{table:wave_method_compare}
\end{table*}

\subsubsection{Results}

The standard deviation of each measured wavelength (for all three methods) from each spectrum is comparable ($\sim 0.01$ {\AA}) to the derived uncertainty in the wavelength shift shown in Table \ref{table:results_2018}. The uncertainty in each measurement is calculated from the standard error in the average of the 4 wavelength ranges used for fitting. This test shows that there are no significant biases introduced by any of the fitting methods. Also, the average wavelength measured for the 3 E1 spectra is consistent across all 3 methods within the measurement uncertainties.



\section{Discussion}

\begin{figure}
\includegraphics[width=\linewidth]{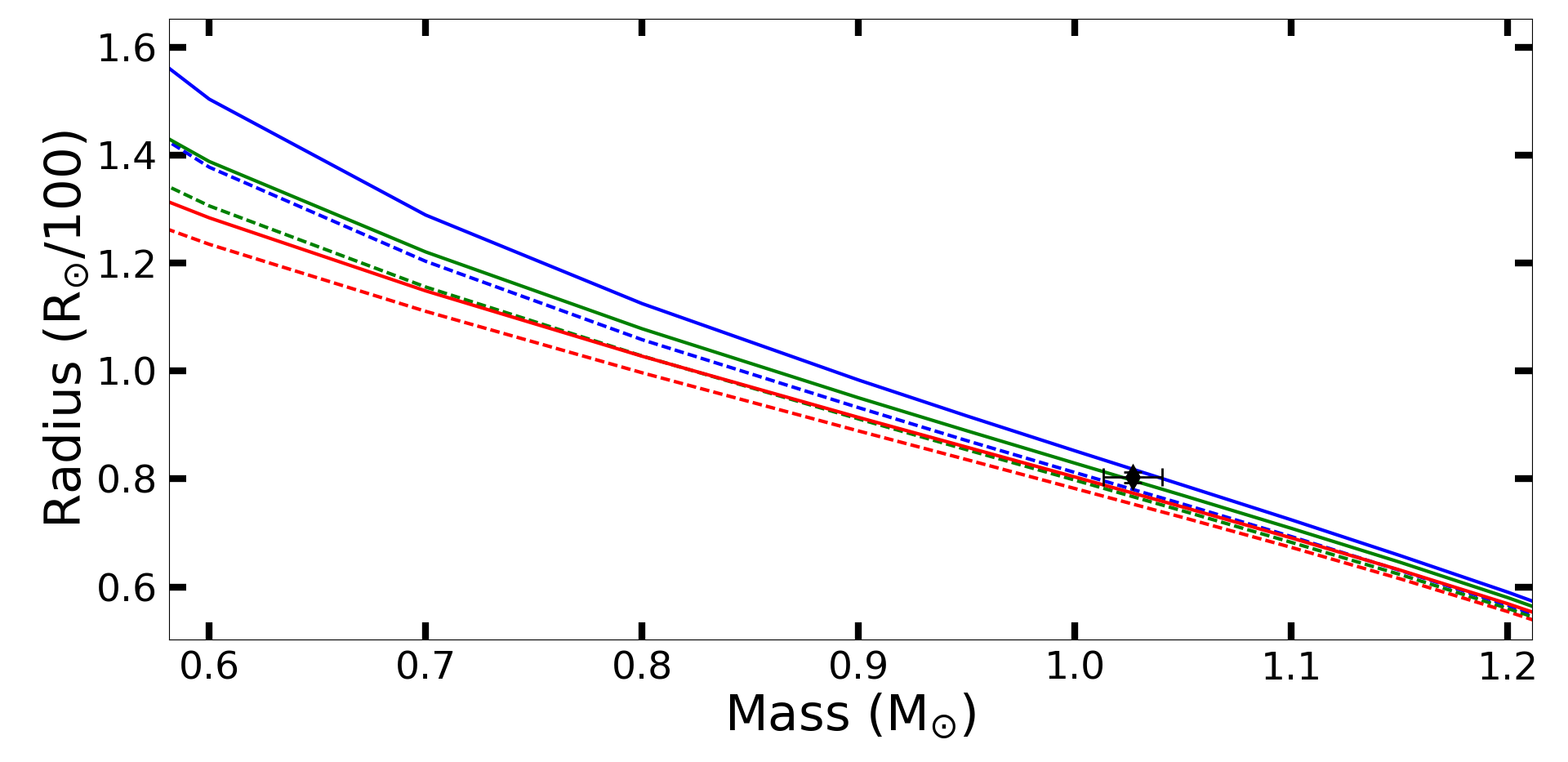}
  \caption{The position of Sirius B on the MRR as measured from the differential gravitational redshift. The theoretical mass-radius relations are from \citep{Fontaine_eta01} and are colour coded according to temperature from 10,000 K (red), 25,922 K (gold), 40,000 K (blue). Dashed lines represent thin H-layer models $(q_{\rm H}=M_{\rm H}/M_{*}=10^{-10}$) and solid lines are thick H-layer models $(q_{\rm H}=M_{\rm H}/M_{*}=10^{-4}$). }
  \label{fig:MRR_diff_Vgr_2018}
\end{figure}

\begin{figure*}
\includegraphics[width=175mm]{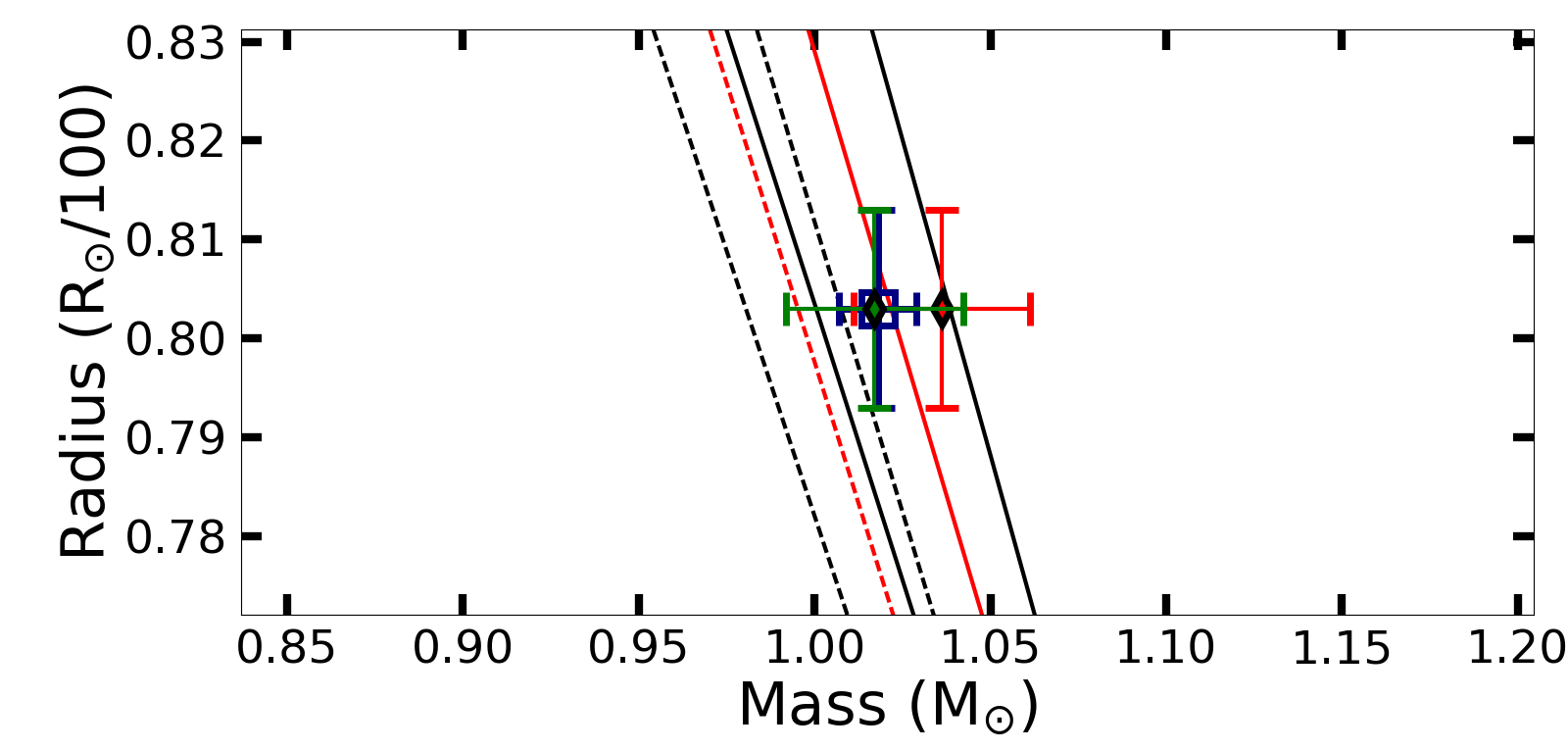}
  \caption{Mass measured from the gravitational red-shift at the E1 position (green diamond on the left) compared to the mass from the dynamical method (dark blue square) \citep{Bond_sirius_17}. The red diamond (right) is the gravitational red-shift mass from the 'centre' position. The solid and dashed lines are \citep{Fontaine_eta01} C/O core mass-radius relations for thick and thin H-layer respectively. The (red) MRR is for a temperature of 25,922 K which is the appropriate T$_{\rm eff}$ for Sirius B according to the spectroscopic fits to the G430 data (Chapter 3). Black lines either side are for 10,000 K (left) and 40,000 K (right). }
  \label{fig:MRR_diff_Vgr_zoom_2018}
\end{figure*}

\subsection{Sirius B and the MRR}
The mass derived from the gravitational redshift measurements can be compared to the theoretical MRR for white dwarfs. The MRR plotted in Fig. \ref{fig:MRR_diff_Vgr_2018} is based on the evolutionary models of \citep{Fontaine_eta01} and includes models for temperatures for 10,000 K (red) to 25,922 K (gold) and 40,000 K (blue). Dashed and solid lines represent thin and thick H-layer respectively. The difference in radius due to temperature declines towards the high mass end so the models converge. 

The mass resulting from the differential redshift measurement is in excellent agreement with the theoretical MRR. Fig. \ref{fig:MRR_diff_Vgr_2018} shows that Sirius B lies directly on the predicted relation. This is clear evidence that white dwarfs follow the expected trend of decreasing radius with increasing mass. 

A close-up of the Sirius B data point is shown in Fig.  \ref{fig:MRR_diff_Vgr_zoom_2018}. Here, the red line indicates a MRR calculated specifically for the temperature of 25,922 K which was found to be the best fit $T_{\rm eff}$ from spectroscopic fitting of the G430L Balmer line spectra. The mass from the E1 (green diamond) and centre (red diamond) data are plotted separately. Both are consistent with the MRR for a 25,922 K C/O core white dwarf. The E1 result is considered to be more reliable because it is based on a larger number of spectra which showed greater consistency than the centre aperture results. For comparison, the dynamical mass from \cite{Bond_sirius_17} is also plotted (blue square). The data points for the E1 redshift and dynamical mass are directly on top of one another despite being obtained using completely different methods. 

It should be noted that a new parallax measurement for Sirius A was included in \textit{Gaia} DR2. This slightly smaller value of 376.68 $\pm$ 0.45 mas gives a radius of 0.808 $\pm$ 0.011 R$\odot/100$ and mass of 1.0246 $\pm$ 0.0268 M$\odot$ based on the E1 gravitational redshift of 80.65 km s$^{-1}$.  However, the accuracy of the new parallax may be affected by the fact that the astrometric solution assumes a linear motion, which is not appropriate for Sirius A due its binary orbit. Currently, our preferred value for the mass is that based on the parallax in \cite{Bond_sirius_17}.

\subsection{The difference between the E1 and Centre results}

It is clear that there is a systematic offset in the wavelengths measured at the E1 position compared to the centre position. The offset is 0.27 \AA\ which would be a significant additional velocity of 12 km s$^{-1}$ if the E1 H$\alpha$ line were to be compared to a laboratory rest wavelength. Two possible causes of this offset were investigated. The first possibility is that it is due to the known issue of the slit angle which is not quite perpendicular to the dispersion direction. This offset to the zero wavelength would normally be around 1-2 km s$^{-1}$ if the target is dithered a small distance along the slit. However, the E1 position places the target at the extreme end of the slit and would have a much larger effect. From the difference between the centre and E1 position, the distance is E1 - Centre = 898 - 512 = 386 pixels in cross-dispersion direction, which gives an offset of 1.19  \AA\ or 54.3 km s$^{-1}$. According to the STIS instrument handbook, the pipeline includes calibration for both the E1 and centre positions so this effect is automatically corrected for. It is also much larger than the offset found so this is unlikely to be the cause.

Another possibility is that the observed offset is a result of the uncertainty of the calibration at the E1 position which is known to be less accurate than the centre position. A study of the relative accuracy of the two positions was carried out by \citealt{Friedman05_stis_inst_report} (STIS Instrument science report). The calibration tests showed that at the centre (row 512) the uncertainty in the mean of the wavelength measurements is 0.05 pixels which at 0.56 \AA\ per pixel is 0.028 \AA. At the E1 position (row 896) the uncertainty increases to 0.2 pixels (0.112\AA). So in terms of velocity the instrumental calibration uncertainty is 1.3 km s$^{-1}$ and 5.1 km s$^{-1}$ at the centre and E1 positions respectively. This could go some way to explaining the observed differences but is not enough to fully account for the offset of around 12 km s$^{-1}$.

\subsection{Evidence of a systematic offset in measured wavelengths}

\begin{table*}

\caption{The velocity ($v_{\rm obs}$) of Sirius A calculated from the shift in its H$\alpha$ line compared to the model rest wavelength. The final velocity is what remains after all known causes of wavelength shift have been removed, which should leave a final velocity of zero. The non-zero values in the final column are therefore a measure of the systematic offset in the measurements.}
\begin{tabular}{ccccc}

\hline

obs ID & Wavelength & $v_{\rm obs}$ & Correction & Final velocity \\

& (\AA) & (km s$^{-1}$) & (km s$^{-1}$) & (km s$^{-1}$) \\
\hline 
odl601030 (E1) & 6564.76$\pm$0.01 & 6.8$\pm$0.5 & 9.55 & 16.3$\pm$0.9\\
odl601040 (E1) & 6564.75$\pm$0.01 & 6.5$\pm$0.7 & 9.55  & 16.0$\pm$1.0\\ 

Average &  &  &  & 16.2\\

&&&&\\
odl601050 (Centre) &  6564.45$\pm$0.02  & -7.3$\pm$0.7 & 9.55 & 2.8$\pm$1.0 \\

odl601060 (Centre) &  6564.46$\pm$0.01  & -6.8$\pm$0.5 & 9.55 & 2.1$\pm$0.9\\ 

Average &  &  &  & 2.5 \\

\hline 
\end{tabular}\label{table:Sirius_A_compared_to_REST}
\end{table*}


From the results obtained with the differential measurements, it is now clear that the velocity measured with respect to the lab rest wavelength in the \cite{Barstow_2017} analysis is systematically too large. It is important to uncover the cause of the offset so that the gravitational redshift method can be applied to other white dwarfs where a convenient reference star may not be available.

The gravitational redshift of Sirius A is known to be less than 1 km s$^{-1}$. Therefore, taking the measured wavelength of Sirius A compared to the lab rest wavelength of H$\alpha$ and correcting for the space motion of Sirius A should yield a zero velocity. Any residual velocity must be an instrumental effect assuming all corrections are applied correctly.

The laboratory rest wavelength used is 6564.6078  \AA\ which is the rest wavelength in a vacuum taken from the NIST Atomic Spectra Database \citep{NIST}. The velocities resulting from the difference between the Sirius A and rest wavelength are listed in Table \ref{table:Sirius_A_compared_to_REST}, column 2. The corrections applied are +8.035  km s$^{-1}$ for the velocity of Sirius A relative to the observer and -0.72 km s$^{-1}$ for the gravitational redshift of Sirius A. The spectra had already been corrected for the \textit{HST} orbital motion as described previously. The resulting velocities are listed in column 5.

The systematic offset in the E1 data could explain the discrepancy in the mass since an offset of 16 km s$^{-1}$ is equivalent to an additional mass of 0.2 M$_{\odot}$. However, it remains unclear if this fully explains the problem as the offset in the centre position is only 2.5 km s$^{-1}$. This is not enough to explain the mass discrepancy in the 2013 spectra which were taken at the centre position.

\subsubsection{Stability of the instrument}

The E1 data for Sirius B are consistent with the average throughout the whole observation showing that there was no scatter introduced due to thermal effects or changes to the instrument settings between the first and last exposure. The only significant effect is the changing velocity due to the orbital motion of \textit{HST}. This is predictable and easily corrected. However, it should be noted that this correction is not included automatically in the pipeline processing.



\section{Conclusions}

The mass of Sirius B as measured by the gravitational redshift method is 1.017 $\pm$ 0.025 M$_{\odot}$. This matches the dynamical mass 1.018 $\pm$ 0.011 M$_{\odot}$ of \cite{Bond_sirius_17} almost exactly. The gravitational redshift mass is consistent with the MRR for a C/O core WD with thick envelope at the temperature of 25,922 K which matches the temperature derived spectroscopically from the G430M spectra. 

This study has shown that the differential method of measuring the wavelength shift is a reliable and accurate method for determining the mass of white dwarfs. The offset between the wavelengths measured from spectra taken at the centre and E1 positions on the CCD, as well as the non-zero shift measurement for Sirius A after all corrections were applied confirms that it is a characteristic of the instrument which caused the offset in the mass measured in previous attempts. Therefore, caution must be used when comparing measured wavelengths from \textit{HST} to lab rest as there appears to be an unexplained offset in the \textit{HST} spectra which is largest for the E1 position ($\sim$ 16 km s$^{-1}$ or 0.35 \AA).

\section*{Acknowledgements}

Special thanks to Blair Porterfield, Joleen Carlberg and the team at STScI for all their help and advice during the phase 2 proposal. We thank the referee for helpful comments. 
SRGJ acknowledges support from the Science and Technology Facilities Council (STFC, UK). 
MAB acknowledges support from the Gaia post-launch support programme of the UK Space Agency and the Leicester Institute for Space and Earth Observation (LISEO).
SLC acknowledges support from LISEO. 
J.B.H. and H.E.B. acknowledge support provided
by NASA through grants from the Space Telescope Science Institute, which is
operated by the Association of Universities for Research in Astronomy, Inc.,
under NASA contract NAS5-26555.

\appendix

\bsp

\label{lastpage}


\begin{thebibliography}{99}

\bibitem[Adams(1925)]{Adams_1925} Adams, W.~S.\ 1925, The Observatory, 48, 337 

\bibitem[Barstow et al.(2005)]{Barstow05} Barstow, M.~A., Bond, H.~E., Holberg, J.~B., et al.\ 2005, MNRAS, 362, 1134 

\bibitem[Barstow et al.(2017)]{Barstow_2017} Barstow, M.~A., Joyce, S., Casewell, S.~L., et al.\ 2017, 20th European White Dwarf Workshop, 509, 383 


\bibitem[Bergeron, Saffer \& Liebert(1992)]{Bergeron_eta92}
{Bergeron, P., Saffer, R. A., \& Liebert, J.} 1992, 
\textit{ApJ}, 394, 228

\bibitem[Bohlin(2014)]{Bohlin_2014AJ_SiriusA} Bohlin, R.~C.\ 2014, AJ, 147, 127 

\bibitem[Bond et al.(2017a)]{Bond_40Eri_17} Bond, H.~E., Bergeron, P., \& B{\'e}dard, A.\ 2017a, ApJ, 848, 16 

\bibitem[Bond et al.(2017b)]{Bond_sirius_17} Bond, H.~E., Schaefer, G.~H., Gilliland, R.~L., et al.\ 2017b, ApJ, 840, 70

\bibitem[Chandrasekhar(1931)]{Chandrasekhar1931} Chandrasekhar, S.\ 1931, MNRAS, 91, 456,``The Highly Collapsed Configurations of a Stellar Mass" 

\bibitem[Einstein.(1916)]{Einstein_1916} Einstein, A.,\ 1916,``The foundation of the general theory of relativity", Annalen der Physik 354, 769-822


\bibitem[Falcon et al.(2010)]{Falcon_2010} Falcon, R.~E., Winget, D.~E., Montgomery, M.~H., \& Williams, K.~A.\ 2010, ApJ, 712, 585 


\bibitem[Falcon et al.(2015)]{Falcon15_ApJ} Falcon, R.~E., Rochau, G.~A., Bailey, J.~E., et al.\ 2015, ApJ, 806, 214

\bibitem[Friedman(2005)]{Friedman05_stis_inst_report} Friedman, S.D.\ 2005, Instrument Science Report STIS 2005-03, ``Wavelength Calibration Accuracy of the
First-Order CCD Modes Using the E1
Aperture" 


\bibitem[Fontaine, Brassard \& Bergeron(2001)]{Fontaine_eta01}
{Fontaine, G., Brassard, P., \& Bergeron, P.} 2001, 
\textit{PASP}, 113, 409

\bibitem[Grabowski \& Halenka(1975)]{Grabowski75} Grabowski, B., \& Halenka, J.\ 1975, A\&A, 45, 159 

\bibitem[Grabowski et al.(1987)]{Grabowski87} Grabowski, B., Halenka, J., \& Madej, J.\ 1987, APJ, 313, 750  

\bibitem[Greenstein et al.(1971)]{Greenstein_1971} Greenstein, J.~L., Oke, J.~B., \& Shipman, H.~L.\ 1971, ApJ, 169, 563 

\bibitem[Greenstein \& Trimble(1967)]{Greenstein_Trim_1967} Greenstein, J.~L., \& Trimble, V.~L.\ 1967, ApJ, 149, 283 

\bibitem[Halenka et al.(2015)]{Halenka2015} Halenka, J., Olchawa, W., Madej, J., \& Grabowski, B.\ 2015, ApJ, 808, 131

\bibitem[Holberg(2010)]{Holberg_2010} Holberg, J.~B.\ 2010, Journal for the History of Astronomy, 41, 41 

\bibitem[Holberg et al.(2012)]{Holberg_2012} Holberg, J.~B., Oswalt, T.~D., \& Barstow, M.~A.\ 2012, AJ, 143, 68 

\bibitem[Joyce et al.(2018)]{Joyce_2018_a} Joyce, S.~R.~G., Barstow, M.~A., Casewell, S.~L., et al.\ 2018, MNRAS, 479, 1612

\bibitem[Kervella et al.(2003)]{Kervella_2003} Kervella, P., Th{\'e}venin, F., Morel, P., Bord{\'e}, P., \& Di Folco, E.\ 2003, A\& A, 408, 681 

\bibitem[Koester(1987)]{Koester_1987} Koester, D.\ 1987, ApJ, 322, 852

\bibitem[Kramida et al.(2018)]{NIST} Kramida, A., Ralchenko, Yu., Reader, J. and NIST ASD Team (20l8). NIST Atomic Spectra Database (version 5.5.2), [Online]. Available: https://physics.nist.gov/asd [Wed Jan 24 2018]. National Institute of Standards and Technology, Gaithersburg, MD.
 

\bibitem[Newville et al.(2014)]{lmfit} Newville, M., Stensitzki, T., Allen, D. B., Ingargiola, A.\ 2014 , LMFIT: Non-Linear Least-Square Minimization and Curve-Fitting for Python, https://doi.org/10.5281/zenodo.11813


\bibitem[Oswalt et al.(1991)]{Oswalt_1991} Oswalt, T.~D., Sion, E.~M., Hammond, G., et al.\ 1991, AJ, 101, 583 


\bibitem[Parsons et al.(2017)]{Parsons_2017} Parsons, S.~G., G{\"a}nsicke, B.~T., Marsh, T.~R., et al.\ 2017, MNRAS, 470, 4473 

\bibitem[Press et al.(1986)]{Press86} Press, W.~H., Flannery, B.~P., \& Teukolsky, S.~A.\ 1986, Numerical recipes. The art of scientific computing, Cambridge: University Press, 1986, 

\bibitem[Reid(1996)]{Reid_1996} Reid, I.~N.\ 1996, AJ, 111, 2000 

\bibitem[Riley(2017)]{Riley_2017} Riley, A.\ 2017, Space Telescope Imaging Spectrograph Instrument Handbook, Cycle 25, Version 16.0

\bibitem[Trimble \& Greenstein(1972)]{Trimble_Greenstein_1972} Trimble, V., \& Greenstein, J.~L.\ 1972, ApJ, 177, 441 

\bibitem[van den Bos(1960)]{van_den_Bos_1960} van den Bos, W.~H.\ 1960, Journal des Observateurs, 43, 145 
 

\bibitem[Wegner(1989)]{Wegner_1989} Wegner, G.\ 1989, IAU Colloq.~114: White Dwarfs, 328, 401 

\bibitem[Wiese \& Kelleher(1971)]{Wiese_Kelleher71} Wiese, W.~L., \& Kelleher, D.~E.\ 1971, ApJ, 166, L59 


\end{thebibliography}
\end{document}